\documentclass[twoside,9pt]{article}
\usepackage{extsizes}
\usepackage[super,sort&compress,comma]{natbib} 
\usepackage[version=3]{mhchem}
\usepackage[left=1.5cm, right=1.5cm, top=1.785cm, bottom=2.0cm]{geometry}
\usepackage{balance}
\usepackage{times,mathptmx}
\usepackage{sectsty}
\usepackage{graphicx} 
\usepackage{lastpage}
\usepackage[format=plain,justification=justified,singlelinecheck=false,font={stretch=1.125,small,sf},labelfont=bf,labelsep=space]{caption}
\usepackage{float}
\usepackage{fancyhdr}
\usepackage{fnpos}
\usepackage[english]{babel}
\usepackage{array}
\usepackage{droidsans}
\usepackage{charter}
\usepackage[T1]{fontenc}
\usepackage[usenames,dvipsnames]{xcolor}
\usepackage{setspace}
\usepackage[compact]{titlesec}

\usepackage{bm}
\usepackage{subcaption}
\usepackage{amssymb}
\graphicspath{{Figures/}}

\newcommand{\anisotropies}{\frac{\Delta\mu}{\bar\mu}-\frac{\Delta\epsilon}{\bar\epsilon}}

\newcommand{\bv}[1]{\boldsymbol{#1}}

\newcommand{\zhat}{\hat{\boldsymbol z}}

\newcommand{\dbyd}[2]{\frac{\partial{#1}}{\partial#2}}

\definecolor{cream}{RGB}{222,217,201}

\begin{document}

\pagestyle{fancy}
\thispagestyle{plain}
\fancypagestyle{plain}{

}

\makeFNbottom
\makeatletter
\renewcommand\LARGE{\@setfontsize\LARGE{15pt}{17}}
\renewcommand\Large{\@setfontsize\Large{12pt}{14}}
\renewcommand\large{\@setfontsize\large{10pt}{12}}
\renewcommand\footnotesize{\@setfontsize\footnotesize{7pt}{10}}
\makeatother

\renewcommand{\thefootnote}{\fnsymbol{footnote}}
\renewcommand\footnoterule{\vspace*{1pt}%
\color{cream}\hrule width 3.5in height 0.4pt \color{black}\vspace*{5pt}} 
\setcounter{secnumdepth}{5}

\makeatletter 
\renewcommand\@biblabel[1]{#1}            
\renewcommand\@makefntext[1]%
{\noindent\makebox[0pt][r]{\@thefnmark\,}#1}
\makeatother 
\renewcommand{\figurename}{\small{Fig.}~}
\sectionfont{\sffamily\Large}
\subsectionfont{\normalsize}
\subsubsectionfont{\bf}
\setstretch{1.125} 
\setlength{\skip\footins}{0.8cm}
\setlength{\footnotesep}{0.25cm}
\setlength{\jot}{10pt}
\titlespacing*{\section}{0pt}{4pt}{4pt}
\titlespacing*{\subsection}{0pt}{15pt}{1pt}

\fancyfoot{}
\fancyfoot[RO]{\footnotesize{\sffamily{1--\pageref{LastPage} ~\textbar  \hspace{2pt}\thepage}}}
\fancyfoot[LE]{\footnotesize{\sffamily{\thepage~\textbar\hspace{3.45cm} 1--\pageref{LastPage}}}}
\fancyhead{}
\renewcommand{\headrulewidth}{0pt} 
\renewcommand{\footrulewidth}{0pt}
\setlength{\arrayrulewidth}{1pt}
\setlength{\columnsep}{6.5mm}
\setlength\bibsep{1pt}

\makeatletter 
\newlength{\figrulesep} 
\setlength{\figrulesep}{0.5\textfloatsep} 

\newcommand{\topfigrule}{\vspace*{-1pt}%
\noindent{\color{cream}\rule[-\figrulesep]{\columnwidth}{1.5pt}} }

\newcommand{\botfigrule}{\vspace*{-2pt}%
\noindent{\color{cream}\rule[\figrulesep]{\columnwidth}{1.5pt}} }

\newcommand{\dblfigrule}{\vspace*{-1pt}%
\noindent{\color{cream}\rule[-\figrulesep]{\textwidth}{1.5pt}} }

\makeatother

\twocolumn[
  \begin{@twocolumnfalse}
\vspace{3cm}
\sffamily
\begin{tabular}{m{4.5cm} p{13.5cm} }

 & \noindent\LARGE{\textbf{Electrokinetic flows in liquid crystal thin films with fixed anchoring}} \\
\vspace{0.3cm} & \vspace{0.3cm} \\

 & \noindent\large{Christopher Conklin$^{\ast}$ and Jorge Vi\~nals} \\ \\

 & \noindent\normalsize{We study ionic and mass transport in a liquid crystalline fluid film in its nematic phase under an applied electrostatic field. Both analytic and numerical solutions are given for some prototypical configurations of interest in electrokinetics: Thin films with spatially nonuniform nematic director that are either periodic or comprise a set of isolated disclinations. We present a quantitative description of the mechanisms inducing spatial charge separation in the nematic, and of the structure and magnitude of the resulting flows. The fundamental solutions for the charge distribution and flow velocities induced by disclinations of topological charge $m=-1/2, 1/2$ and $1$ are given. These solutions allow the analysis of several designer flows, such as  \lq\lq pusher" flows created by three colinear disclinations, the flow induced by an immersed spherical particle (equivalent to an $m=1$ defect) and its accompanying $m=-1$ hyperbolic hedgehog defect, and the mechanism behind nonlinear ionic mobilities when the imposed field is perpendicular to the line joining the defects.
} \\

\end{tabular}

 \end{@twocolumnfalse} \vspace{0.6cm}

  ]

\renewcommand*\rmdefault{bch}\normalfont\upshape
\rmfamily
\section*{}
\vspace{-1cm}


\footnotetext{\textit{School of Physics and Astronomy, University of Minnesota, 116 Church St. SE, Minneapolis, MN 55455, USA. E-mail:conk0044@umn.edu}}




\onecolumn

\section{Introduction}
\label{sec:introduction}

Electrokinetic effects such as electrophoresis and electroosmosis are important tools for particle, second phase, or fluid transport in a wide variety of engineering, soft matter, and biological systems \cite{re:morgan03,re:squires04,re:bazant09}. The fundamental requirement for the existence of electrokinetic phenomena in a multicomponent fluid is the ability to spatially separate electrostatic charge. Once separated, relative charge motion can be driven by an applied electric field, which in turn drives fluid flow or particle transport within the system. In isotropic electrolytes, charge separation occurs due to electric double layer formation on fluid-solid surface boundaries, when the boundaries are sufficiently polarizable. Our focus here instead is on liquid crystal dispersions in which the fluid matrix is a liquid crystal in its nematic phase. Electrokinetic effects can be induced in the bulk by an applied electric field when the director configuration is not uniform.

The physical interactions that are present in colloidal dispersions in which the matrix is a liquid crystal have been reviewed by Stark\cite{re:stark01}. The basic phenomenology of these dispersions differs from that of suspensions involving simple fluids in two ways: First, a specific anchoring orientation of the liquid crystal molecule on the surface of a suspended particle leads to a long range distortion of the nematic director field, to an excess elastic energy in the matrix, and ultimately to elastic interactions among suspended particles. Second, a given anchoring orientation of the liquid crystal molecule on the surface of a suspended particle can induce a topological defect in the nematic director field. Depending on the far field conditions, a new defect or set of defects need nucleate in the matrix to compensate for the topological charge associated with the suspended particle. Therefore, the analysis of equilibrium configurations, transport, and hydrodynamic flows in nematic colloidal suspensions requires in general the simultaneous consideration of both suspended particles and associated topological defects in the suspending fluid. Novel, single particle effects investigated to date include levitation and unidirectional drift near container walls \cite{re:pishnyak07}, or anisotropic and both super- and sub-diffusive Brownian motion at time scales that correspond to the relaxation times of director distortions around the particles \cite{re:turiv13}. Collective effects addressed include the formation of chain like structures \cite{re:poulin97}, and particle aggregation driven by the topological defects in the matrix \cite{re:araki06}. Of particular interest for applications of liquid crystal suspensions is the fact that fluid trajectories (and hence the motion of solid or immiscible liquid inclusions) can be controlled through pre-designed director configurations, as demonstrated by circular orbiting of colloids \cite{re:lavrentovich10}, or disordered mixing flows. Dynamic control has also been demonstrated by light irradiation \cite{re:hernandez-navarro14}.

In classical electrokinetics there is a linear relationship between the applied electric field and any induced velocities. Hence direct current fields (DC) are required to produce systematic transport, which is then limited to the time needed for the ions in the bulk electrolyte to move towards the electrodes and screen the applied field. Another general consequence of linearity is that the flow is irrotational, $\nabla \times \mathbf{v} = 0$, thus limiting the applicability of the configuration (e.g., no vortical flows for mixing applications). Transport by DC fields is also affected by other problems, such as undesirable electrochemical reactions at surfaces, relatively low velocities, and poor control of surface charge \cite{re:lavrentovich15}. These difficulties can be overcome when charge separation in the suspension itself is induced by the applied field so that the resulting velocities are quadratic in the field intensity. Systematic transport can the be induced by an alternating current (AC) applied field instead. In the liquid crystalline matrices considered, charge separation can be achieved in bulk due to either anisotropic dielectric permittivity or anisotropic mobilities of dissolved ionic impurities, both a function of the local nematic orientation.

As a first step in the analysis of the complex phenomenology that can arise in nematic suspensions, we focus here on approximately two dimensional configurations in which a nonuniform nematic director field is held fixed. Such a configuration can be realized experimentally by placing a nematic thin film between two bounding surfaces on which a specific director orientation is specified through lithographic treatment of the surfaces \cite{re:peng15}. We begin by considering a simple periodic configuration with wavevector perpendicular to the electrostatic field. This configuration is formally equivalent to that found in electrohydrodynamic convection studies \cite{re:kaiser92}, and can be largely solved analytically. We then focus on two-dimensional configuration that contain a single disclination in the imposed director field with topological charge $m = -1/2, 1/2, 1$. We analytically obtain the corresponding fundamental solutions for the induced charge density and velocity field. A numerical finite element method is then presented that can treat more general configurations. The code is first validated against both periodic and disclination induced transport. We then numerically examine a few configurations comprising an ensemble of disclinations and show that they can be used for flow design. For example, when the electric field is parallel to the line joining two or more disclinations with zero total topological charge, the resulting flow can serve to push or pull mass in pre specified directions. If the electric field is perpendicular to that line, defect induced charge asymmetries result in transverse flow, and hence in transverse nonlinear mobilities. We finally discuss the effect of combined anisotropic dielectric permittivity of the liquid crystal and anisotropic ionic mobility, and show that both effects can either reinforce or counteract each other depending on the signs of the anisotropies. For a binary mixture of liquid crystals, this effect can lead to complete flow reversal depending simply on the composition of the fluid.


\section{Model description}
\label{sec:model}

Electrokinetic phenomena in liquid crystals are qualitatively different than in isotropic fluids, as spatial charge separation can be induced in bulk without the need of any solid surfaces, and it follows  from spatially varying director field \cite{re:lavrentovich10,re:lazo13,re:lazo14}. Two different microscopic mechanisms leading to electrokinetic phenomena are considered here that follow from the anisotropic nature of the liquid crystal molecule: anisotropy in the dielectric permittivity of the molecule along the directions parallel and perpendicular to the director, and anisotropic mobility of dissolved ionic impurities \cite{re:helfrich69}. Liquid crystals always contain some amount of ionic impurities that are present as residuals in chemical synthesis, through adsorption from adjacent media (such as alignment layers), and through injection of charges from electrodes in contact with the liquid crystal. Electrokinetic phenomena in liquid crystals with positive dielectric anisotropy have been investigated experimentally in the mixture of E7 with MLC7026-000 \cite{re:lavrentovich10}. Changing the mixture composition allowed the exploration of a range of relative dielectric anisotropies from 13.8 to as low as 0.03. Negative dielectric anisotropy has also been investigated by Lazo, et. al. \cite{re:lazo13} More recently, electrokinetic phenomena have been studied in a mixture of E7 and MLC7026-000 of a specific composition so that the mixture has negligible dielectric anisotropy, $| \Delta \epsilon | \le 10^{-3}$ \cite{re:peng15}. In this latter study, electrokinetic transport is entirely attributable to dissolved ionic impurities. 

Our analysis is based on the Leslie-Ericksen model \cite{re:degennes93,re:stewart04} in which the local orientation of the nematic is described by the director field $\hat{\bm{n}}(\mathbf{r})$. In this description, topological defects in the nematic appear as singularities in the director field that need to be handled carefully. However, although the boundary conditions used approximate singular nematic configurations, the solutions for charge distributions and velocity fields within the film are themselves not singular, and hence difficulties associated with short scale divergences in the solution do not explicitly arise.

\subsection{Transport model}

Consider $k = 1, \ldots, N$ ionic species of charge $e z_k$, where $e$ is the elementary (positive) charge and $z_{k}$ an integer, immersed in a liquid crystalline fluid (neutrally charged). We assume that the masses of the ionic species are small compared to the masses of the liquid crystal molecules. The liquid crystal is in its nematic phase (exhibiting long range orientational order, but no positional order), characterized by the local nematic director $\hat{\bm{n}}(\mathbf{r})$, a unit vector describing the average orientation of the molecules in a small element of volume at $\mathbf{r}$. Nematic order requires invariance under $\hat{\bm{n}} \rightarrow -\hat{\bm{n}}$. We write continuity for the $k$ species as $\partial_{t} c_{k} + \nabla \cdot \mathbf{J}_{k} = 0$, where $c_{k}$ is the concentration of species $k$ and $\mathbf{J}_{k}$ its number density flux. Using standard irreversible thermodynamics for electrolyte solutions \cite{re:degroot84}, we decompose the flux $\mathbf{J}_{k}$ into a reactive component including advection of a local element of volume at the barycentric velocity $\bv{v}$ (which includes the liquid crystal), and a dissipative contribution arising from species diffusion and drift induced by the electrostatic field $\bv{E} = - \nabla \Phi$; thus $\mathbf{J}_k = c_k \bv v - \mathbf{D} \cdot\nabla c_k -  c_k z_k \bm{\mu} \cdot \nabla\Phi$. The quantities $\bv{D}$ and $\bm{\mu}$ are the diffusivity and ionic mobility tensors respectively, which will be assumed to be anisotropic and depend on the local orientation of the liquid crystalline molecule. They are also assumed to obey  Einstein's relation $\bv{D} = \frac{k_{B}T}{e} \bm{\mu}$. Given the ratios of masses between the ionic species and the liquid crystalline matrix, we will approximate $\bv{v}$ in what follows by the velocity of the liquid crystal alone. By using mass continuity, we arrive at the equation governing the evolution of the concentration of species $k$,
\begin{equation}
\dbyd{c_k}{t}+ \nabla\cdot(\bv v c_k)=\nabla\cdot\left(\bv{D}\cdot\nabla c_k + c_k z_k \bm{\mu} \cdot\nabla\Phi \right)
\label{eq:concentration}
\end{equation}

The mobility tensor $\bv\mu$ is assumed to be anisotropic and to depend on the local orientation of the nematic \cite{re:helfrich69}. In Cartesian components, $\mu_{ij} = \mu_\perp\delta_{ij}+\Delta \mu \; n_i n_j$ where $\delta_{ij}$ is the Kroenecker delta, and we define $\Delta\mu = \mu_{\parallel}-\mu_{\perp}$, where $\mu_\parallel$ and $\mu_\perp$ are the ionic mobilities parallel and perpendicular to $\hat{\bm{n}}$, respectively. If $n_{0}$ is the equilibrium concentration of carriers in the system, we further define the electrical conductivity tensor $\sigma_{ij} = e n_{0} \mu_{ij}$. 

In the time scales of interest, the system is assumed to be in electrostatic equilibrium, so that the total electrostatic potential in the medium $\Phi$ satisfies
\begin{equation}
\label{eq:Poisson}
-\epsilon_0\nabla \cdot (\bv\epsilon\cdot\nabla \Phi)=
\sum^{N}_{k=1} ez_kc_k
\end{equation}
Although the liquid crystal molecules are not charged, they are polarizable \cite{re:degennes93}. The nematic is assumed to be a linear dielectric medium, with dielectric tensor $\epsilon_{ij} = \epsilon_{\perp}\delta_{ij} + \Delta\epsilon \; n_i n_j$, with $\Delta\epsilon = \epsilon_\parallel-\epsilon_\perp$, where $\epsilon_\parallel$ and $\epsilon_\perp$ are the dielectric constants parallel and perpendicular to $\hat{\bf{n}}$, respectively. 

In terms of momentum conservation, we assume that the total mass density and velocity $\bv{v}$ of an element of volume can be well approximated by those of the liquid crystal. The liquid crystal is incompressible, $\nabla\cdot\bv v=0$, and flow is overdamped. Typical velocities considered are $v \sim 10 \mu m/s$ or less, and characteristic length scales of the flows we model $L$ are on the order of hundreds of $\mu m$. With liquid crystal densities $\rho \sim 10^3 kg/m^{3}$ and viscosities $\eta \sim  0.1 kg/(m s)$ \cite{re:degennes93,re:kleman03} one estimates the Reynolds number $Re = \rho v L/\eta$ to be $Re \sim  10^{-5} - 10^{-4}$; hence inertia can be neglected. Momentum balance then reduces to the balance between stresses and the body force exerted by the ionic species in a field,
\begin{equation}
0 =\nabla \cdot \mathbf{T} - \sum_{k=1}^N e z_kc_k\nabla\Phi
\label{eq:momentum_conservation}
\end{equation}
In Cartersian components, the stress tensor is $T_{ij}=-p \delta_{ij}+T_{ij}^e + \tilde T_{ij}$, where $p$ is the pressure and $\bv{T^e}$ is the elastic stress,
\begin{equation}
T_{ij}^e = -\dbyd{f}{(\partial_j n_k)}\dbyd{n_k}{x_i}
\end{equation}
where $f$ is the Oseen-Frank elastic free energy density\cite{re:degennes93}. The viscous stress $\bv{\tilde{T}}$ is,
\begin{equation}
\tilde T_{ij} = \alpha_1 n_i n_j n_k n_l A_{kl} + \alpha_2 N_i n_j + \alpha_3 n_iN_j + \alpha_4 A_{ij}+\alpha_5 n_j A_{ik}n_k+\alpha_6 n_i A_{jk} n_k
\label{eq:leslie}
\end{equation}
with $N_i = \dot n_i - \Omega_{ij}n_j$, and $A_{ij}=\frac12\left(\dbyd{v_j}{x_i}+\dbyd{ v_i}{x_j}\right)$
and $\Omega_{ij} = \frac12\left(\dbyd{v_j}{x_i}-\dbyd{v_i}{x_j}\right)$ the symmetric and antisymmetric parts of the velocity gradient tensor. The coefficients $\alpha_{i}$ are the Leslie viscosities \cite{re:stewart04}, and $\dot{n}_i=\partial n/\partial t + (\bv v\cdot\nabla)n_i$.

Finally, we consider angular momentum conservation, which defines the dynamics of the director $\bv{\hat n}$. For a nematic in an electric field, the balance of torques yields:\cite{re:degennes93}
\begin{equation}\label{eq:directorDynamics}
    \bv{\hat n}\times{\bv h^0} -\bv{\hat n}\times{\bv h^v}+\epsilon_0\Delta\epsilon(\bv{\hat n}\cdot\bv E)(\bv{\hat n}\times\bv E)=0,
\end{equation}
$$h^0_i = -\dbyd{f}{n_i}+\dbyd{}{x_j}\dbyd{f}{(\partial_j x_i)},$$
$$h^v_i = (\alpha_3-\alpha_2) N_i+(\alpha_3+\alpha_2)A_{ij}n_j$$

Equations (\ref{eq:concentration}) through (\ref{eq:directorDynamics}) are the complete set of governing equations together with the incompressibility condition. A more complete derivation of this transport model has been given by Calderer, et al. \cite{re:calderer16}. These equations are highly nonlinear, and can produce a variety of complex behavior. In particular, the nonlinear coupling of Eq. (\ref{eq:momentum_conservation}) and Eq. (\ref{eq:directorDynamics}) alone can produce a variety of effects such as tumbling, wagging, and chaotic motion\cite{re:rienacker02}. In order to investigate electrokinetic effects only, we choose to study systems in which elastic torques on the nematic dominate over viscous and dielectric torques, or equivalently, systems in which the Ericksen number, $\text{Er} = \eta v L/K$, and the ratio $\epsilon_0\Delta\epsilon E^2 L^2/K$ are both small, so induced reorientation of the nematic and flexoelectric effects may be neglected. Therefore the director orientation in a thin film is assumed to be uniform in the $z$ direction (the thickness direction), but can have a specified $(x,y)$ dependence, determined by the alignment imposed by the bounding surfaces. Consistent with this approximation, we also assume that the ionic charge distribution and velocity fields are two dimensional.

To further simplify the problem, we consider only two species of charge $z_{1} = 1$ and $z_{2} = -1$, and the system as a whole is assumed electrically neutral. The fluid is assumed under a uniform and oscillatory electrostatic field along the $x$ direction, $\mathbf{E} = E_{0}\cos(\omega t) \hat{\bm{x}}$.


$$
$$

\subsection{Periodic anchoring}
\label{sec:periodic}


To illustrate the mechanisms of electrokinetic flow and to validate our numerical analysis, we begin by investigating a domain with a periodically anchored director, $\hat{\bm n}(y)=(\cos(qy),\sin(qy))$. This is a simple case that can be solved analytically to obtain the scalings of the change density and flow velocity in terms of the physical parameters. Concentrations, velocities, and induced electric fields do not depend on $x$. Furthermore, by symmetry $v_y=0$. For two ionic species we define $\Delta c = c_1-c_2$ and $C = c_1+c_2$ and write Eq. (\ref{eq:concentration}) as
\begin{equation}\label{eq:C_periodic}
    \dbyd{C}{t} = \dbyd{}{y}\left(D_{yy}\dbyd{C}{y}+\Delta c \mu_{yi}\dbyd{\Phi}{x_i}\right)
\end{equation}
\begin{equation}\label{eq:dc_periodic}
    \dbyd{\Delta c}{t} = \dbyd{}{y}\left(D_{yy}\dbyd{\Delta c}{y}+C\mu_{yi}\dbyd{\Phi}{x_i}\right)
\end{equation}

\begin{figure}[h]
\centering
\begin{subfigure}{0.4\textwidth}
\includegraphics[width=\linewidth]{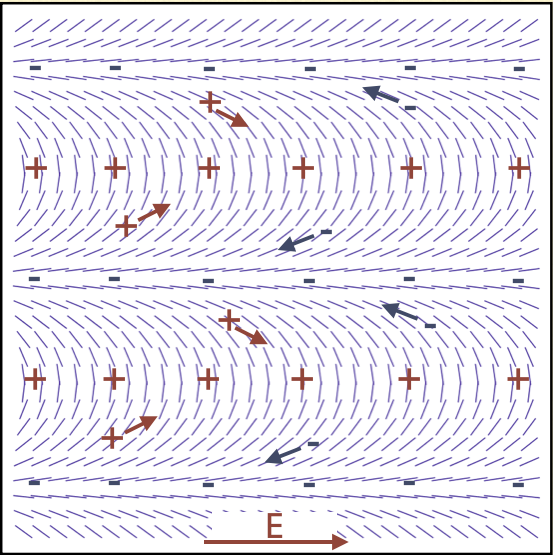}
\caption{\label{fig:ChargeSeparationCartoon}}
\end{subfigure}
\begin{subfigure}{0.4\textwidth}
\includegraphics[width=\linewidth]{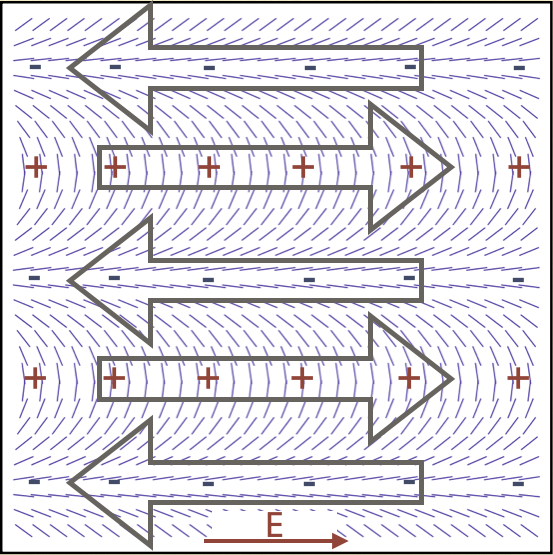}
\caption{\label{fig:FlowCartoon}}
\end{subfigure}
\caption{(\subref{fig:ChargeSeparationCartoon}) Ionic mobility anisotropy leads positive and negative ions to collect in different regions of the cell under an applied electric field. (\subref{fig:FlowCartoon}) Regions with high concentration of positive ions will flow in the direction of the electric field, while regions of high negative concentration will flow opposite to the electric field.}
\label{fig:MobilityCartoon}
\end{figure}

Under the applied field, anisotropic mobility leads to charge separation through the drift term in the right-hand side of Eq. (\ref{eq:dc_periodic}), and is schematically illustrated in Figure \ref{fig:ChargeSeparationCartoon} (the director is shown as thin lines in the figure). With an electric field in the $+\hat{\bm{x}}$ direction, positive ions move to the right. Since the mobility is higher parallel to $\hat{\bm{n}}$, positive ions drift towards regions where $\hat{\bm{n}}$ is parallel to the $y$ direction. Negative ions on the other hand drift to the left towards regions where $\hat{\bm{n}}$ is parallel to the $x$ direction.

\begin{figure*}[h]
\begin{subfigure}{0.33\textwidth}
\includegraphics[width=\linewidth]{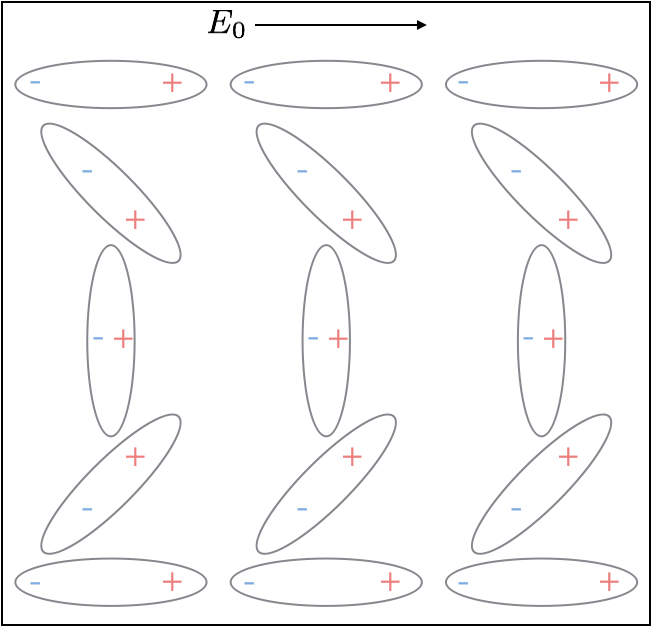}
\caption{\label{fig:polarizationcartoon1}}
\end{subfigure}
\begin{subfigure}{0.33\textwidth}
\includegraphics[width=\linewidth]{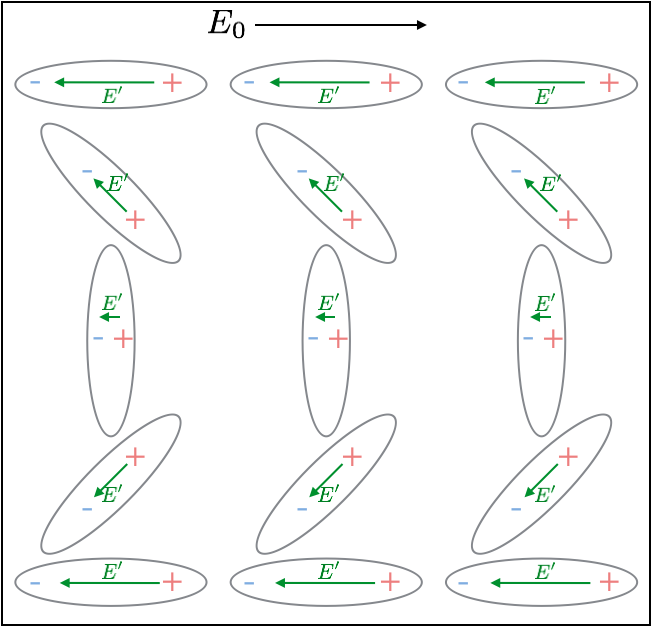}
\caption{\label{fig:polarizationcartoon2}}
\end{subfigure}
\begin{subfigure}{0.33\textwidth}
\includegraphics[width=\linewidth]{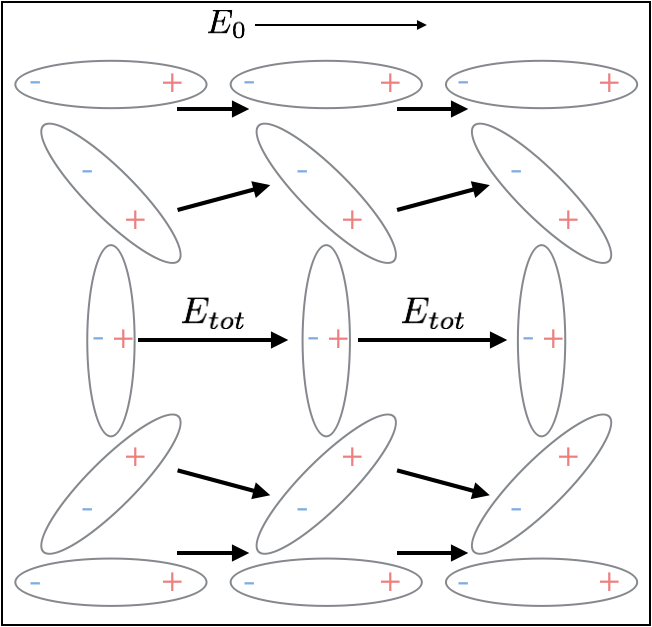}
\caption{\label{fig:polarizationcartoon3}}
\end{subfigure}
\begin{subfigure}{0.33\textwidth}
\includegraphics[width=\linewidth]{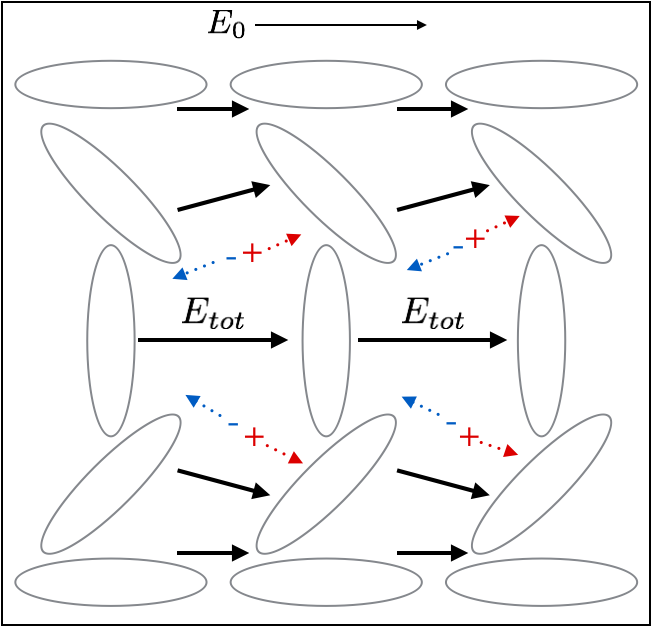}
\caption{\label{fig:polarizationcartoon4}}
\end{subfigure}
\begin{subfigure}{0.33\textwidth}
\includegraphics[width=\linewidth]{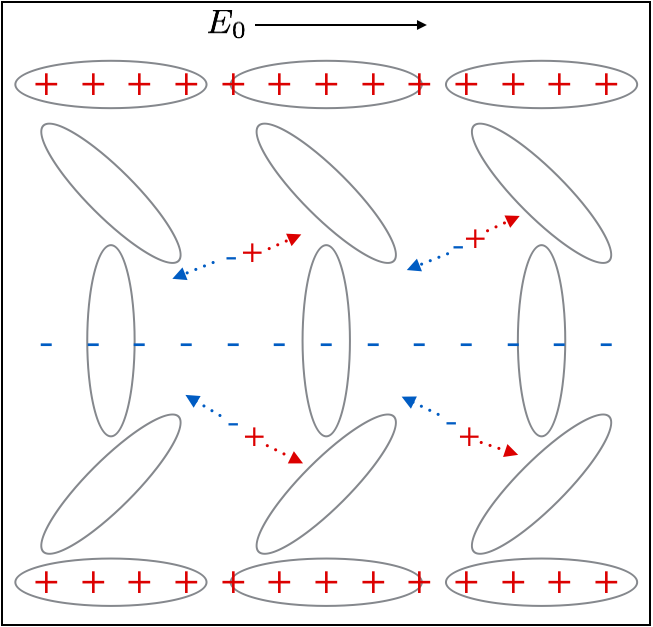}
\caption{\label{fig:polarizationcartoon5}}
\end{subfigure}
\begin{subfigure}{0.33\textwidth}
\includegraphics[width=\linewidth]{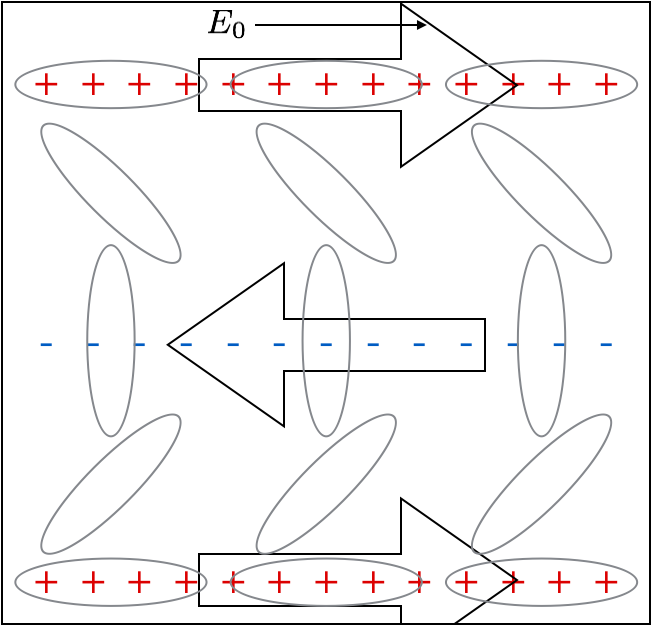}
\caption{\label{fig:polarizationcartoon6}}
\end{subfigure}
\caption{(\subref{fig:polarizationcartoon1}) Dielectric anisotropy leads to polarization which varies as a function of director orientation. (\subref{fig:polarizationcartoon2})-(\subref{fig:polarizationcartoon3}) Nematic polarization generates nonuniform electric fields in response to the applied field, creating an overall nonuniform electric field. (\subref{fig:polarizationcartoon4})-(\subref{fig:polarizationcartoon5}) The nonuniform electric field causes positive and negative ions to collect in different regions of the fluid. Note the sign of the charges in a given region is opposite that of Fig. \ref{fig:ChargeSeparationCartoon}. (\subref{fig:polarizationcartoon6}) Regions of positive and negative charge density drive nematic flow in opposite directions.}
\label{fig:DielectricCartoon}
\end{figure*}

Charge separation due to dielectric permittivity anisotropy also occurs via the drift term in the right-hand side of Eq. (\ref{eq:dc_periodic}) and its coupling with Eq. (\ref{eq:Poisson}); Figure \ref{fig:DielectricCartoon} illustrates this mechanism. If the dielectric permittivity is higher parallel to the director ($\Delta \epsilon > 0$), polarization will reduce the electric field more strongly when it is parallel to the director than perpendicular. Thus an applied field in the $+\hat{\bm x}$ direction will generate a total electric field with sources where $\hat{\bm n}$ is parallel to the $y$ axis and sinks where $\hat{\bm n}$ is parallel to $x$ axis, as shown in Fig. \ref{fig:polarizationcartoon3}. Therefore the concentration of positive ions will be high in regions where $\hat{\bm n}$ is parallel to $\hat{\bm x}$, while the concentration of negative ions will be higher in regions where $\hat{\bm n}$ is parallel to $\hat{\bm y}$, as shown in Fig. \ref{fig:polarizationcartoon5}. 

After charge separation has taken place, fluid elements experience an electrostatic body force through the last term on the right-hand-side of Eq. (\ref{eq:momentum_conservation})  as indicated schematically in Figs. \ref{fig:FlowCartoon} and \ref{fig:polarizationcartoon6}. Note that the net charge density in a given region due to dielectric anisotropy has a sign opposite to that of charge separation due to mobility anisotropy, thus implying the two mechanisms will counteract each other. Importantly, when the electric field polarity is inverted, charge separation is also reversed, but the body force direction is unchanged, hence leading to systematic flow, even under an AC field.

For both mechanisms, charge separation saturates by species diffusion and the additional transverse field $- \partial \Phi/\partial y$ induced by charge separation itself. Both effects lead to different characteristic times for saturation. If saturation is due to the induced transverse field, the characteristic time scale is $\tau_{\rho} = \bar\epsilon \epsilon_{0}/(n_{0}e \bar\mu)$. On the other hand, the mass diffusion time is $\tau_{D} = (\bar Dq^{2})^{-1}$, with $q$ set by the director patterning scale. For typical liquid crystal parameters, $\bar\epsilon\sim 10$, $\bar\mu\sim 10^{-9}~m^2/(Vs)$, $n_0\sim10^{19}~m^{-3}$, $2\pi/q\sim 100\mu m$, and at room temperature, We find $\tau_\rho\sim 0.1~s$ and $\tau_D\sim 10~s$, hence diffusive saturation of charge separation is negligible for pattern scales on the order of $ 100~\mu m$ or smaller. Furthermore, both characteristic times need to be compared to the frequency of the applied field $\omega$. From Eq. (\ref{eq:dc_periodic}) when $\omega \gg 1/\tau_\rho$, $\Delta c$ will be out of phase with the imposed field by $\pi/2$. In the limit $\omega \rightarrow 0$, on the other hand, the imposed field and $\Delta c$ will be in phase. Therefore, since the body force on the liquid crystal medium, Eq. (\ref{eq:momentum_conservation}), involves the product of the charge and the applied field, we expect systematic transport to occur only for sufficiently low frequencies, $\epsilon_{0} \bar{\epsilon} \omega/ n_{0}e \bar{\mu} \ll 1$.

Charge separation can also induce variations in the total concentration $C$ by the last term on the right-hand side of Eq. (\ref{eq:C_periodic}). We estimate the size of this term relative to $\partial C/\partial t$ by first using Poisson's equation to estimate the size of $\Delta c$: $\Delta c\sim \epsilon_0\bar\epsilon q E_0/e$. Then if we assume the scale of $C$ is $C\sim n_0$, we estimate
\begin{equation}
\frac{\left|\dbyd{}{y}\left(\Delta c\mu_{yi}[\partial{\Phi}/\partial{x_i}]\right)\right|}{\left|\partial{C}/\partial{t}\right|}\sim \frac{(q E_0)^2\bar\epsilon\epsilon_0\bar\mu}{\omega e n_0}.
\end{equation}
This ratio is very small, and variations in $C$ are negigible so that $C\approx n_{0}$, independent of time. Under these conditions, and by defining the charge density $\rho = e \Delta c$, Eq. (\ref{eq:dc_periodic}) becomes simply
\begin{equation}
\label{eq:nc}
\dbyd{\rho}{t} =  \dbyd{}{y}\left(\sigma_{yj} \dbyd{\Phi}{x_j} \right),
\end{equation}
where we recall  and $\sigma_{ij} = \sigma_\perp\delta_{ij}+(\sigma_\parallel-\sigma_\perp)n_in_j= n_{0} e\mu_{ij}$. Equations (\ref{eq:Poisson}) and (\ref{eq:nc}) can be solved to yield
\begin{equation}
\label{eq:chargeDensity}
\rho(y,t) = \left(-\frac{\Delta\sigma}{\bar\sigma}+\frac{\Delta\epsilon}{\bar\epsilon}\right)\epsilon_0\bar\epsilon E_0\dbyd{}{y}\left[\frac{\bar\sigma\cos(\omega t - \delta(y))\sin(2 q y)}{2\sqrt{[\sigma_{yy}(y)]^2+\left[\omega\epsilon_0\epsilon_{yy}(y)\right]^2}}\right], \quad 
\tan\delta(y) = \frac{\omega\epsilon_0\epsilon_{yy}(y)}{\sigma_{yy}(y)}
\end{equation}
where $\Delta\sigma = \sigma_\parallel-\sigma_\perp$ and $\bar\sigma = (\sigma_\parallel+\sigma_\perp)/2$. 

We note that the charge density vanishes linearly with anisotropies $\Delta\sigma$ and $\Delta\epsilon$, is linear in the applied field, and oscillates with frequency $\omega$ at a phase shift $\delta(y)$ from the applied field. As anticipated, charge separation due to mobility and dielectric anisotropy are of opposite signs in a given region. If the director orientation variation along $y$ is small, $\bv n \approx (1,\varphi)$, then $\rho = \left(-\frac{\Delta\sigma}{\bar\sigma}+\frac{\Delta\epsilon}{\bar\epsilon}\right)\epsilon_0\bar\epsilon E_x \dbyd{\varphi}{y}$, as already reported by Lazo, et. al\cite{re:lazo14}.

By assuming no variation along $x$, Eq. (\ref{eq:momentum_conservation}) reduces to $ \dbyd{}{y}\tilde T_{xy} +\rho E_x = 0$, which can be integrated explcitly,
\begin{equation}\label{eq:PeriodicStressTensor}
\tilde T_{xy} = \left(\frac{\Delta\sigma}{\bar\sigma}-\frac{\Delta\epsilon}{\bar\epsilon}\right)\frac{\epsilon_0\bar\epsilon\bar\sigma E_0^2\cos(\omega t)\cos(\omega t-\delta(y))\sin(2 q y)}{2\sqrt{[\sigma_{yy}(y)]^2 + [\omega\epsilon_0\epsilon_{yy}(y)]^2}}
\end{equation}
At high frequencies, $\omega \gg \sigma_{yy}/(\epsilon_{0} \epsilon_{yy})$, $\delta = \pi/2$, and the stress oscillates around zero with a small average. At low frequencies, $\delta \approx \pi$, and the resulting stress has a non zero average, hence leading to systematic flow.

By using $\nabla\cdot\bv v=0$, the assumption of variation in $y$ only, and fact that $\hat{\bv n}$ is fixed, one finds
\begin{equation}
\label{eq:gradv}
\dbyd{v_x}{y} = \left(\frac{\Delta\sigma}{\bar\sigma}-\frac{\Delta\epsilon}{\bar\epsilon}\right)\frac{\epsilon_0\bar\epsilon\bar\sigma E_0^2\sin(2qy)\cos(\omega t)\cos(\omega t-\delta(y))}{2\sqrt{[\sigma_{yy}(y)]^2 + [\omega\epsilon_0\epsilon_{yy}(y)]^2}\left[\frac{\alpha_1}{2}\sin^2(2 q y) + (\alpha_2+\alpha_3)\cos(2qy)+\alpha_3+\alpha_4+\alpha_5  \right]}
\end{equation}
This equation can be integrated exactly. Like the charge density, the velocity is proportional do the difference in relative mobility and dielectric anisotropy. The velocity magnitude is proportional to the square of the electric field, and the phase shift $\delta(y)$ indicates that as the driving frequency increases, the systematic part of the velocity approaches zero.


We mention that the analytic expression for the flow field suggests a method for experimentally determining the Miezowicz viscosities of the liquid crystal. At $y=0$, the velocity is parallel to the director and hence
\begin{equation}\label{y0Momentum}
\frac12(\alpha_3+\alpha_4+\alpha_6)\left[\partial_y^2 v_x\right ]_{y=0} =\left(-\frac{\Delta\sigma}{\bar\sigma}+\frac{\Delta\epsilon}{\bar\epsilon}\right) \frac{\bar\epsilon\epsilon_0\bar\sigma E_0^2\cos(\omega t)\cos(\omega t-\delta_{\perp})}{\sqrt{\sigma_\perp^2+(\omega\epsilon_\perp\epsilon_0)^2}}, \; \tan \delta_\perp = \frac{\omega\epsilon_0\epsilon_\perp}{\sigma_\perp}
\end{equation}
whereas at $y=\pi/(2q)$, we find,
\begin{equation}\label{yPiHalfMomentum}
\frac12(-\alpha_2+\alpha_4+\alpha_5)\left[\partial_y^2 v_x\right ]_{y=\frac{\pi}{2q}} = \left(-\frac{\Delta\sigma}{\bar\sigma}+\frac{\Delta\epsilon}{\bar\epsilon}\right) \frac{\bar\epsilon\epsilon_0\bar\sigma E_0^2\cos(\omega t)\cos(\omega t-\delta_{\parallel})}{\sqrt{\sigma_\parallel^2+(\omega\epsilon_\parallel\epsilon_0)^2}},\; 
\tan \delta_\parallel = \frac{\omega\epsilon_0\epsilon_\parallel}{\sigma_\parallel}
\end{equation}
The Leslie viscosity combinations in the left hand sides of Eqs. (\ref{y0Momentum}) and (\ref{yPiHalfMomentum}) precisely correspond to the Miezowicz viscosities, $\eta_\parallel=\eta^b=\frac12(\alpha_3+\alpha_4+\alpha_6)$ and $\eta_\perp = \eta^c = \frac12(-\alpha_2+\alpha_4+\alpha_5)$. Therefore one can use the experimentally determined flow fields near $y=0$ and $y=\pi/2$ to obtain $\eta_\parallel$ and $\eta_\perp$.

\subsection{Numerical Model}

We have developed a finite element code to solve the Eqs. (\ref{eq:concentration}) through (\ref{eq:leslie}) on a square domain with a prescribed director pattern $\hat{\bv n}(\bv x)$. The finite element method and other numerical details are given in Appendix \ref{sec:numeric}. We have used Eqs. (\ref{eq:chargeDensity}) and (\ref{eq:gradv}) to validate the code for a periodic director pattern in a region of parameters for which the approximate equations discussed hold, a region which is consistent with the experimental parameters described in Appendix \ref{sec:numeric}. Figure \ref{fig:PeriodicComparison} shows good agreement between the numerical solutions and Eqs.(\ref{eq:chargeDensity}) and (\ref{eq:gradv}). 

\begin{figure}[h]
\begin{subfigure}{0.5\textwidth}
\includegraphics[width=0.9\linewidth]{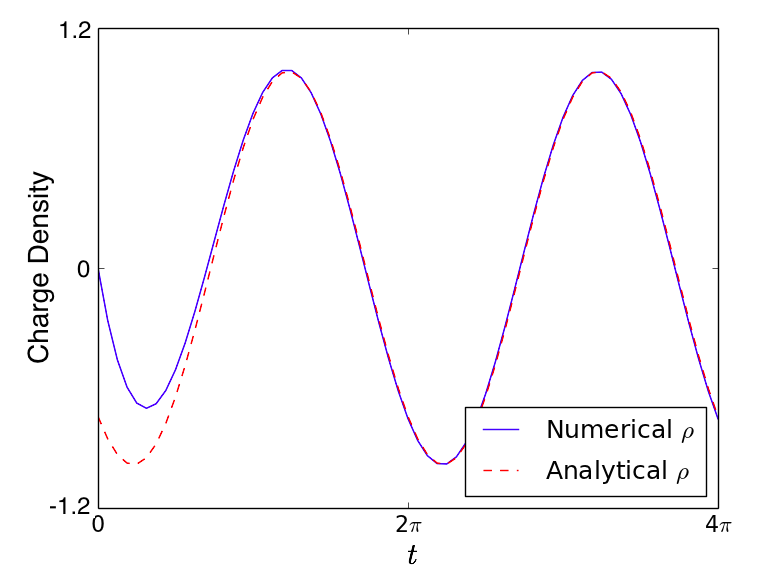}
\caption{\label{fig:ChargeDensityTime}}
\end{subfigure}
\begin{subfigure}{0.5\textwidth}
\includegraphics[width=0.9\linewidth]{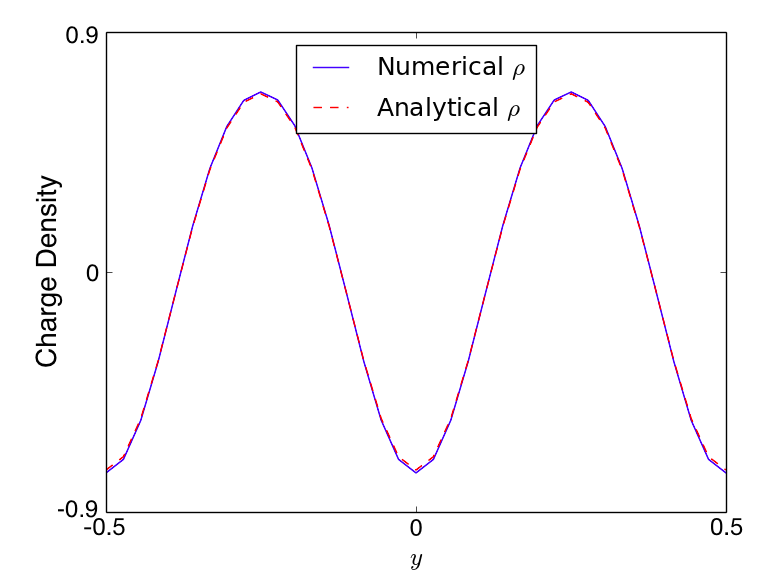}
\caption{\label{fig:ChargeDensitySpace}}
\end{subfigure}
\begin{subfigure}{0.5\textwidth}
\includegraphics[width=0.9\textwidth]{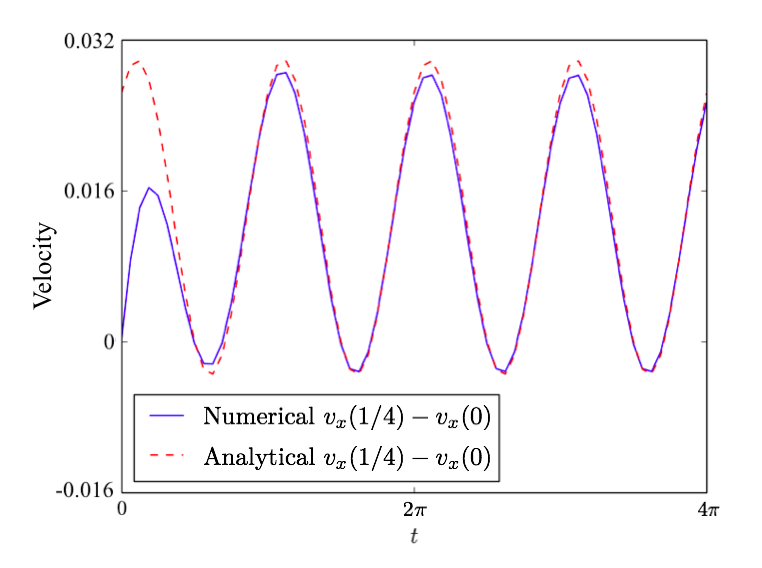}
\caption{\label{fig:VelocityTime}}
\end{subfigure}
\begin{subfigure}{0.5\textwidth}
\includegraphics[width=0.9\textwidth]{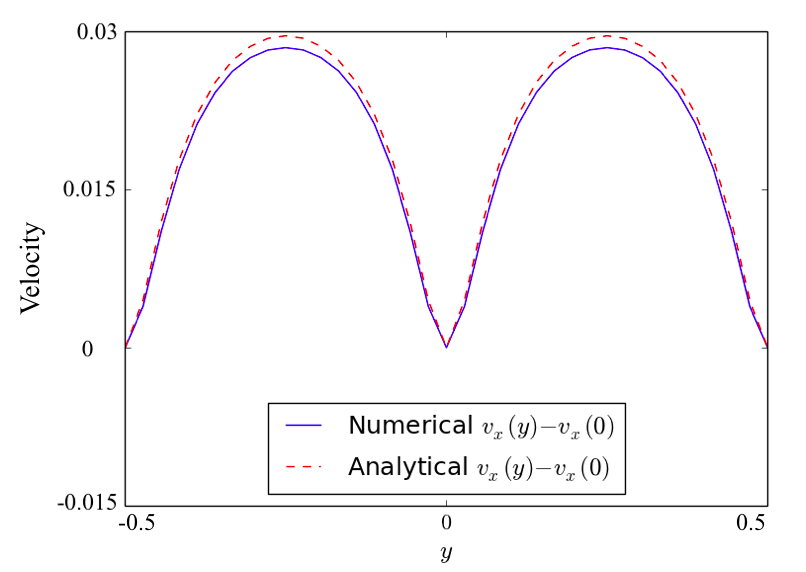}
\caption{\label{fig:VelocitySpace}}
\end{subfigure}
\caption{Analytic and numerical solutions in dimensionless units for charge density and velocity for periodic director patterning. (\subref{fig:ChargeDensityTime}) Charge density as a function of time at $y=0$. (\subref{fig:ChargeDensitySpace}) Charge density as a function of $y$ at $t=2\pi$. (\subref{fig:VelocityTime}) Velocity difference between $y=1/4$ and $y=0$ as a function of time. (\subref{fig:VelocitySpace}) Velocity as a function of $y$ relative to velocity at $y=0$, at $t=2\pi$.}
\label{fig:PeriodicComparison}
\end{figure}

\begin{figure}[h]
\begin{subfigure}{0.29\textwidth}
\includegraphics[width=0.9\linewidth]{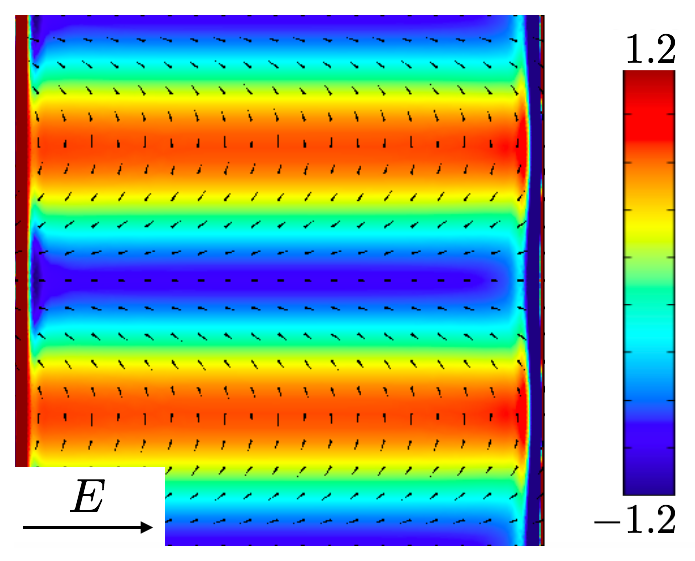}
\caption{\label{fig:PeriodicCharge}}
\end{subfigure}
\begin{subfigure}{0.32\textwidth}
\includegraphics[width=0.9\linewidth]{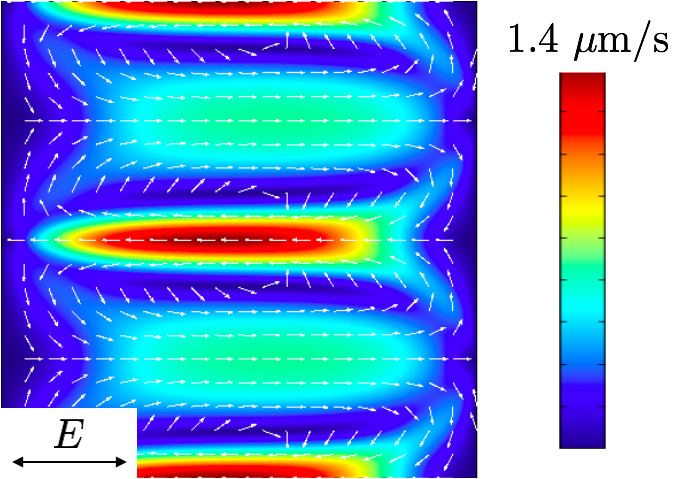}
\caption{\label{fig:PeriodicVelocityNum}}
\end{subfigure}
\begin{subfigure}{0.36\textwidth}
\includegraphics[width=\linewidth]{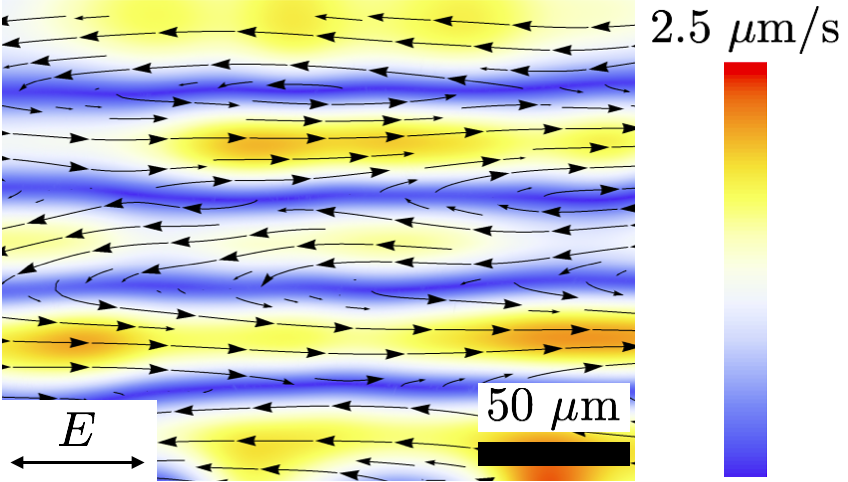}
\caption{\label{fig:PeriodicVelocityExperiment}}
\end{subfigure}
\caption{Numerical and experimental results for a periodically anchored director. (\subref{fig:PeriodicCharge}) Numerical charge density in dimensionless units at $t=2\pi$. (\subref{fig:PeriodicVelocityNum}) Velocity field averaged over a period of the electric field. (\subref{fig:PeriodicVelocityExperiment}) Experimental velocity for a periodic director pattern obtained by Particle Image Velocimetry by time averaging over the locations of tracer particles \cite{re:peng15}.}
\label{fig:PeriodicNumericalSurface}
\end{figure}

To further validate our model, we compare our numerical results with the experiments of  of Peng, et. al.\cite{re:peng15} While our numerical calculations for the periodic pattern use periodic boundary conditions along $y$ and no slip boundary conditions on $x=\pm 1/2$, the experiments involve a small patterned sub-region within a larger cell with uniform top and bottom boundaries. Therefore open boundary conditions would be a closer representation of the experiments. As seen in Fig. \ref{fig:PeriodicCharge}, the boundary layers near the ends of the computational domain are much smaller than the domain, and we have verified that the results presented are independent of system size. Nevertheless, Fig. \ref{fig:PeriodicNumericalSurface} shows good agreement between our numerical results for the parameters listed in Table \ref{tab:params} and the experiments of Peng, et. al.\cite{re:peng15} (Fig. \ref{fig:PeriodicVelocityExperiment}).

\begin{figure}[h]
\begin{subfigure}{0.32\textwidth}
\includegraphics[width=\linewidth]{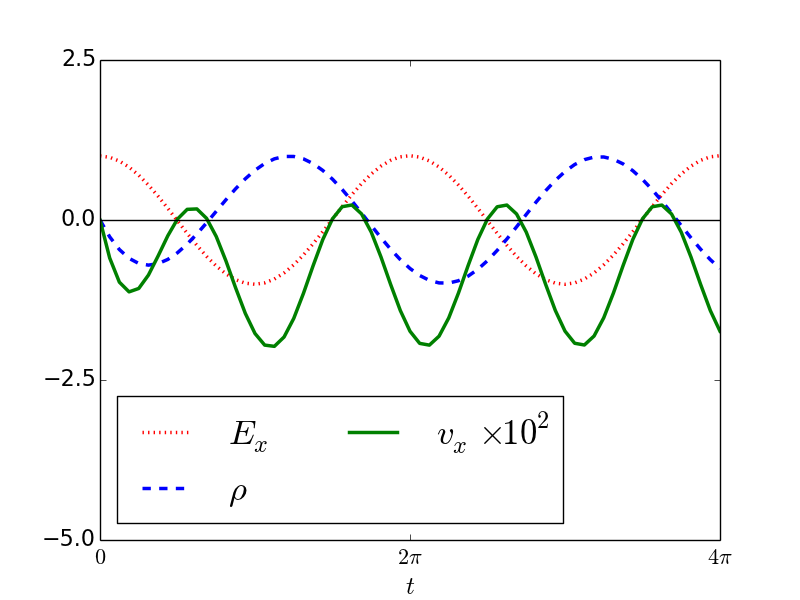}
\caption{$\Delta\epsilon/\bar\epsilon=0,~\Delta\sigma/\bar\sigma=0.34$}
\end{subfigure}
\begin{subfigure}{0.32\textwidth}
\includegraphics[width=\linewidth]{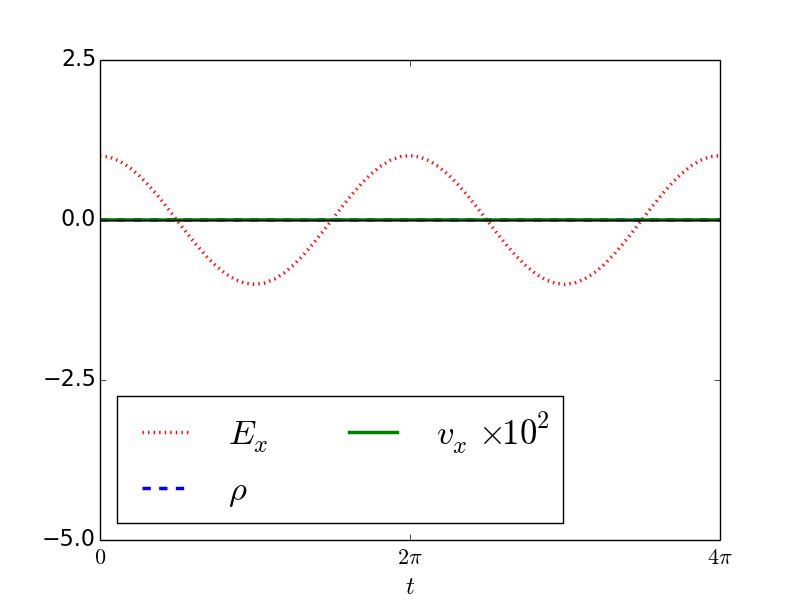}
\caption{$\Delta\epsilon/\bar\epsilon=0.34,~\Delta\sigma/\bar\sigma=0.34$}
\end{subfigure}
\begin{subfigure}{0.32\textwidth}
\includegraphics[width=\linewidth]{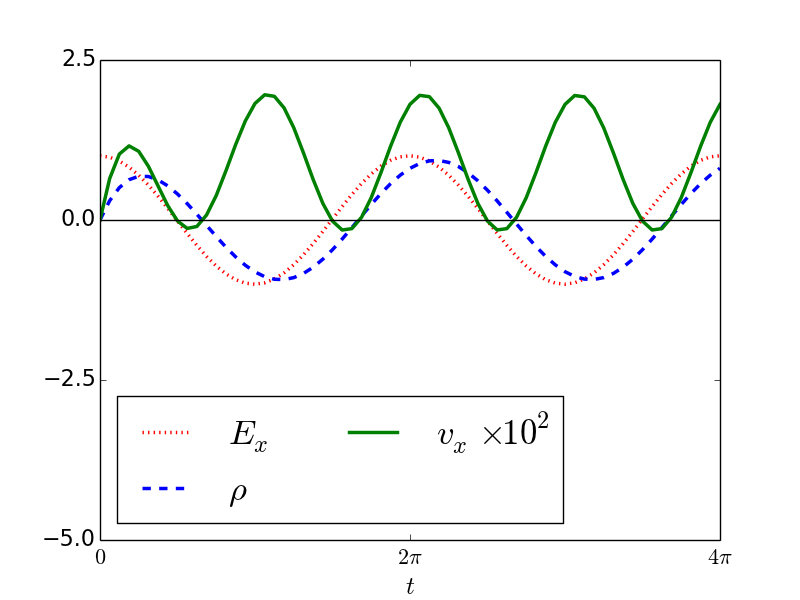}
\caption{$\Delta\epsilon/\bar\epsilon=0.34,~\Delta\sigma/\bar\sigma=0$}
\end{subfigure}
\caption{Numerical solution showing the applied electric field, induced charge density, and $x$ component of the velocity in dimensionless units at $(0,0)$ for periodic anchoring. The velocity direction changes when the quantity $\Delta\epsilon/\bar\epsilon-\Delta\sigma/\bar\sigma$ changes sign.}
\label{fig:AnisotropyComparison}
\end{figure}

One particular consequence of Eqs. (\ref{eq:PeriodicStressTensor}) and (\ref{eq:gradv}) is that flow can be reversed or completely stopped by simply changing the signs of the anisotropies in dielectric permittivity or ionic mobility. Flow reversals have been observed in isotropic electrolytes, but the mechanism is not yet understood \cite{re:bazant09}. In the nematic case, reversals arise from competing charge separation fluxes as described in Figs. \ref{fig:MobilityCartoon} and \ref{fig:DielectricCartoon}. Flow reversals are illustrated by the numerical solution of the full set of governing equations as shown in Fig. \ref{fig:AnisotropyComparison} where we consider the cases of $\Delta\epsilon/\bar\epsilon-\Delta\sigma/\bar\sigma<0$, $\Delta\epsilon/\bar\epsilon-\Delta\sigma/\bar\sigma=0$, and $\Delta\epsilon/\bar\epsilon-\Delta\sigma/\bar\sigma>0$, with all other parameters constant.

\section{Charge separation and flow induced by disclinations}
\label{sec:disclinations}

Having validated our model and numerical code for the simple case of periodic patterning, we now look to study more complex patterns. Lithographic surface patterning offers the opportunity of tailoring flow fields in nematics for specific applications, for example, to engineer flows in microfluidic channels, or to effect immersed particle motion or species separation. We focus here on the case of isolated disclination patterns as they can be used as building blocks for complex designer flow fields.

Consider a two-dimensional configuration with a fixed nematic director orientation $\hat{\bm n}(\bm r)=(\cos\theta(\bm r),\sin\theta(\bm r))$, where $\theta(\bv r)$ is the angle between the director and the $x$ axis. A disclination of topological charge $m$ is given by $\theta(\bv r) = m\phi$, where $\phi$ is the polar angle. We again define the charge density $\rho = e(c_1-c_2)$ and the total concentration $C = c_1+c_2$. We scale lengths by the system size $L$, time by applied field frequency $\omega$, total concentration $C$ by $n_0$, and electric potential by $E_0 L$. The Leslie viscosities $\alpha_i$ are scaled by their average, $\eta$. For the ranges of parameters of the experiments we are interested in, we anticipate the scales for the charge density and velocity to be similar to those in Sec \ref{sec:periodic}; therefore we scale charge density by $\epsilon_0\bar\epsilon L^{-1} E_0$ and velocity by $\epsilon_0\bar\epsilon L \eta^{-1} E_0^2$. The resulting equations are still complex, so we focus first on the limit of small anisotropy, and further expand the dimensionless variables $\rho, C, \mathbf{v}$ and $\Phi$ (we use the same notation as for dimensionless variables. From here onwards, all variables are assumed to be dimensionless) in powers of $\Delta \mu/ \bar\mu$ and $\Delta\epsilon/\bar\epsilon$, both assumed small and of the same order. 

At zero-th order in $\Delta$, the equations correspond to a purely isotropic medium with $c_{1,0}=c_{2,0} = 1/2$ ($\rho_0=0$, $C_0 = 1$), and $\bv v_{0} = 0$. Zero charge density also implies from (\ref{eq:Poisson}) that $\nabla^{2}\Phi_0 = 0$ which we take equal to the imposed field $\Phi_0 = -x\cos(t)$.  

At first order, Eq. (\ref{eq:concentration}) becomes
\begin{equation}
\label{eq:C_scaled}
\Omega \dbyd{C_1}{t} = \gamma\nabla^{2}C_1 - Y^2\cos(t) \dbyd{\rho_1}{x} 
\end{equation}
\begin{equation}
\label{eq:rho_scaled}
\Omega\dbyd{\rho_1}{t} = \gamma\nabla^{2}\rho_1 -\rho_1-\left(\anisotropies\right)\frac{\cos[(2m-1)\phi]\cos(t)}{r}-\cos(t)\dbyd{C_1}{x}.
\end{equation}
There is a driving term in the right hand side of Eq. (\ref{eq:rho_scaled}) which explicitly shows $\anisotropies$ as a coefficient multiplying the order one angular factor. We have further defined $\Omega = \omega \tau_\rho$ (as in Sec. \ref{sec:periodic}, $\tau_\rho =\epsilon_0\bar\epsilon/ (e n_0 \bar\mu)$), $\gamma = \tau_\rho \bar D/L^2$, and $Y=\epsilon_0\bar\epsilon E_0/(L e n_0)$, the charge density relative to the total ionic concentration. Note that  $\gamma$ can be also be written as $\gamma = \lambda_D^2/L^2$, where $\lambda_D = \sqrt{\epsilon_0\bar\epsilon k_B T/(e^2 n_0)}$ is the Debye length. For typical values as given in Appendix \ref{sec:numeric}, $\lambda_D \sim 10^{-6} m$, whereas cell sizes are on the order of $L \sim 10^{-4}-10^{-3} m$, so $\gamma \sim 10^{-6}-10^{-4}$. Thus the term proportional to $\gamma$ in Eqs. (\ref{eq:C_scaled}) and (\ref{eq:rho_scaled}) is a singular perturbation, negligible away from the disclination core, but important within a distance on the order of the Debye length from the core.

We write $\rho$ and $C$ as Fourier series,

\begin{equation}\label{eq:fourier}
\rho_1(\bv r,t) = \sum_n a_n(\bv r) e^{int},\; C = \sum_n b_n(\bv r)e^{int}
\end{equation}
Inserting Eq. (\ref{eq:fourier}) into Eqs. (\ref{eq:C_scaled}) and (\ref{eq:rho_scaled}) yields

\begin{equation}\label{eq:C_n}
\Omega i n b_n = \gamma\nabla^2 b_n - \frac{Y^2}{2}\dbyd{}{x}(a_{n-1}+a_{n+1})
\end{equation}
\begin{equation}\label{eq:rho_n}
\Omega i n a_n = \gamma\nabla^2a_n -a_n - \frac{1}{2}\dbyd{}{x}(b_{n-1}+b_{n+1})-\left(\anisotropies\right)\frac{m\cos[(2m-1)\phi]}{2 r}(\delta_{n,1}+\delta_{n,-1})
\end{equation}
where $\delta_{m,n}$ is Kronecker delta.

We assume a system in which $Y^2/(4\gamma\sqrt{1+\Omega^2})\ll 1$, and we show in App. \ref{sec:term_neglection} that this assumption implies $C$ and $\rho$ decouple. Thus for $n=1$, Eq. (\ref{eq:rho_n}) becomes

\begin{equation}\label{eq:rho_simp}
\gamma\nabla^2 a_1(r,\phi) -(1+\Omega i)a_1(r,\phi) = \left(\anisotropies\right)\frac{m \cos[(2m-1)\phi]}{2r}
\end{equation}
Define $\xi=r/\sqrt{\gamma/(1+i\Omega)}$; then the solution to Eq. (\ref{eq:rho_simp}) is
\begin{equation}\label{eq:a_1}
a_1(\xi,\phi) = \left(\anisotropies\right)\frac{m\cos[(2m-1)\phi]}{2\sqrt{\gamma(1+\Omega i)}} f(\xi)
\end{equation}
We note that the angular dependence of charge density is a function of the topological charge $m$, varying as $\cos[(2m-1)\phi]$, and the radial dependence $f(\xi)$ solves,
\begin{equation}\label{eq:f_xi}
f''(\xi)+\frac{1}{\xi}f'(\xi)-\left(\frac{(2m-1)^2}{\xi^2}+1\right)f(\xi) = \frac{1}{\xi}
\end{equation}
Note the radial dependence of the charge density is in general dependent on $m$. The homogeneous solutions to this differential equation are modified Bessel functions, $I_{|2m-1|}(\xi),K_{|2m-1|}(\xi)$. The particular solutions can be found through variation of parameters, but may also be written in a simpler form for certain vales of $m$, which we discuss below.

For $m=1/2$, the solution to Eq. (\ref{eq:f_xi}) bounded for all $\xi$ is
\begin{equation}\label{eq:f1_full}
f(\xi)=\frac{\pi}{2}\left(L_0(\xi)-I_0(\xi)\right)
\end{equation}
where $L_0(\xi)$ is the modified Struve function of order zero. For $\xi\ll1$, $I_0(\xi)\rightarrow 1$ and $L_0(\xi)\rightarrow\frac2\pi \xi$, so 
\begin{equation}
f(\xi\rightarrow0)\rightarrow\xi-\frac{\pi}{2}.
\end{equation}
The charge density is finite as $r \rightarrow 0$, but it can be large, scaling as $\gamma^{-1/2}$.
For $\xi\gg 1$, given the asymptotic relation \cite{re:abramowitz72},
\begin{equation}\label{eq:StruveLargeX}
L_\alpha(\xi)=I_{-\alpha}(\xi)-\frac{\left(\frac \xi2\right)^{\alpha-1}}{\Gamma\left(\alpha+\frac12\right)\sqrt{\pi}}+\mathcal{O}(\xi^{\alpha-3})
\end{equation}
one has
\begin{equation}\label{eq:f_large}
f(\xi\rightarrow\infty)\rightarrow -\frac{1}{\xi}.
\end{equation}

For $m=1$, the solution to Eq. (\ref{eq:f_xi}) bounded for all $\xi$ is 
\begin{equation}
f(\xi)=K_1(\xi)-\frac{1}{\xi}.
\end{equation}
Note $f(\xi)\rightarrow 0$ as $r\rightarrow 0$ and $f(\xi)$ matches Eq. (\ref{eq:f_large}) for $\xi\gg 1$.

For $m \neq 1/2,1$, the solution to Eq. (\ref{eq:f_xi}) can be obtained by variation of parameters,
\begin{equation}\label{eq:f_var}
f(\xi) = I_{|2m-1|}(\xi)\int^\xi K_{|2m-1|}(\xi')d\xi' - K_{|2m-1|}(\xi)\int^\xi I_{|2m-1|}(\xi')d\xi'
\end{equation}
The asymptotic behavior of $f(\xi)$ at long distances can be found by recalling the asymptotic expansions for $I_k(\xi), K_k(\xi)$ for large $\xi$:
\begin{equation}
I_k(\xi) = \frac{e^{\xi}}{\sqrt{2\pi\xi}}\left(1+\mathcal{O}\left(\frac{1}{\xi}\right)\right),
\end{equation}
\begin{equation}
K_k(\xi) = \sqrt{\frac{\pi}{2\xi}}e^{-\xi}\left(1+\mathcal{O}\left(\frac{1}{\xi}\right)\right).
\end{equation}
For $\xi\gg 1$, using integration by parts,
\begin{equation}
\int^\xi I_k(\xi')d\xi' = \frac{e^{\xi}}{\sqrt{2\pi\xi}}+\int^\xi\mathcal{O}(\xi'^{-3/2}e^{\xi'})d\xi',
\end{equation}
\begin{equation}
\int^\xi K_k(\xi')d\xi'= -\sqrt{\frac{\pi}{2\xi}}e^{-\xi}+\int^\xi\mathcal{O}(\xi'^{-3/2}e^{-\xi'})d\xi'.
\end{equation}
Therefore for $\xi\gg 1$, we find
\begin{equation}
f(\xi)\rightarrow -\frac{1}{\xi}.
\end{equation}
For $\xi\ll 1$, note for integral $k>1$,
\begin{equation}
I_k(\xi) = \frac{1}{k!}\left(\frac{\xi}{2}\right)^k(1+\mathcal{O}(\xi^2))
\end{equation}
\begin{equation}
K_k(\xi)=\frac{(k-1)!}{2}\left(\frac{2}{\xi}\right)^k(1+\mathcal{O}(\xi^2))
\end{equation}
Using these expansions in (\ref{eq:f_var}) we find, to leading order in $\xi$ for $\xi\ll 1$,
\begin{equation}
f(\xi) \sim \frac{\xi}{|2m-1|([2m-1]^2-1)}
\end{equation}
So $a_1$ linearly approaches zero as $\xi\rightarrow 0$.

Using Eq. (\ref{eq:fourier}), we may write the charge density in its complete form
\begin{equation}\label{eq:rho_sol}
\rho(\xi,\phi,t) = \left(\anisotropies\right)\frac{m \cos[(2m-1)\phi]}{2\sqrt{\gamma}(1+\Omega^2)^{\frac{1}{4}}}[e^{i(t-\delta/2)} f(\xi)+e^{-i(t-\delta/2)}f^*(\xi)]
\end{equation}
where $\tan\delta = \Omega$. Note that while the form of  $f(\xi)$ depends on $m$, for all values of $m$ the charge density is linear in $\xi$ near the defect core (approaching zero for $m\neq1/2$), and decays as $\xi^{-1}$ far from the core. In particular, for $\xi\gg 1$ Eq. (\ref{eq:rho_sol}) can be approximated as
\begin{equation}\label{eq:rho_far}
\rho(r,\phi,t) = -\left(\anisotropies\right)\frac{m \cos(t-\delta)}{\sqrt{1+\Omega^2}}\frac{\cos[(2m-1)\phi]}{r}
\end{equation}

\begin{figure*}[h]
\centering
\begin{subfigure}{0.3\textwidth}
\includegraphics[width=\linewidth]{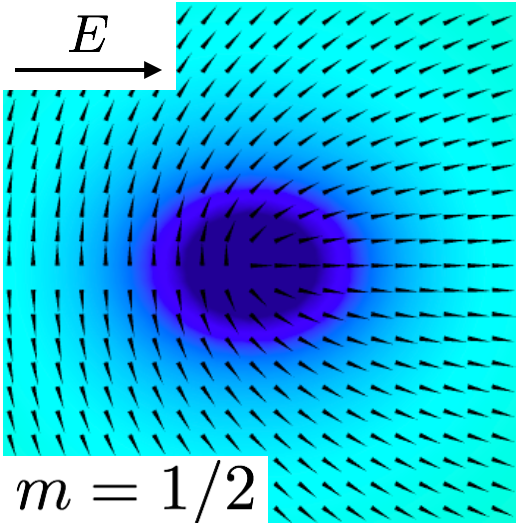}
\caption{\label{fig:pHalfSurf}}
\end{subfigure}
\begin{subfigure}{0.3\textwidth}
\includegraphics[width=\linewidth]{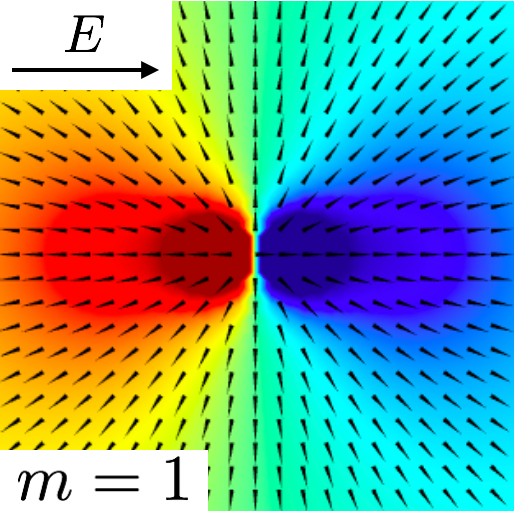}
\caption{\label{fig:p1Surf}}
\end{subfigure}
\begin{subfigure}{0.3\textwidth}
\includegraphics[width=\linewidth]{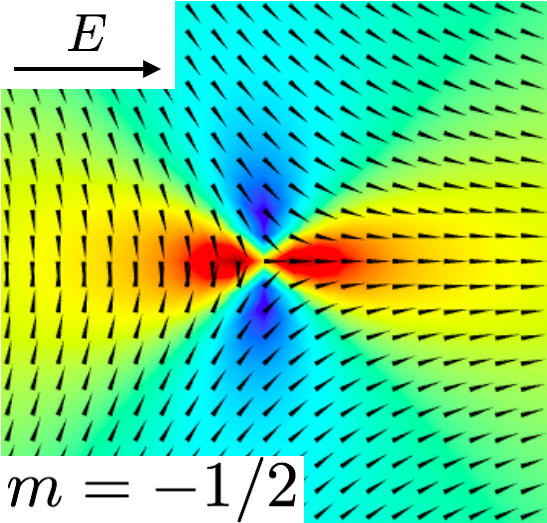}
\caption{\label{fig:mHalfSurf}}
\end{subfigure}
\begin{subfigure}{0.06\textwidth}
\includegraphics[width=\linewidth]{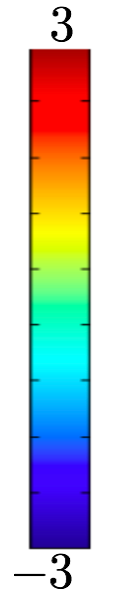}
\caption*{}
\end{subfigure}
\begin{subfigure}{0.3\textwidth}
\includegraphics[width=\linewidth]{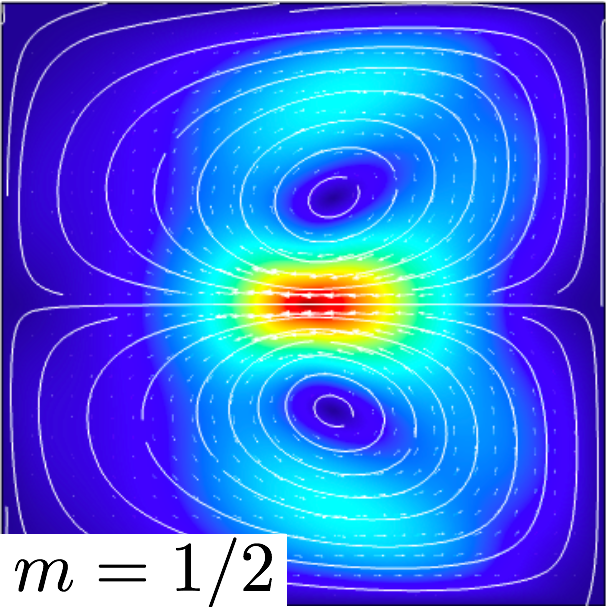}
\caption{\label{fig:pHalfV}}
\end{subfigure}
\begin{subfigure}{0.3\textwidth}
\includegraphics[width=\linewidth]{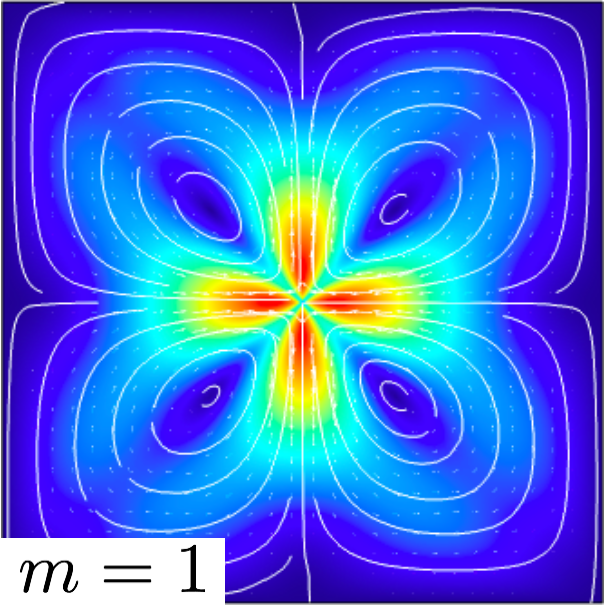}
\caption{\label{fig:p1V}}
\end{subfigure}
\begin{subfigure}{0.3\textwidth}
\includegraphics[width=\linewidth]{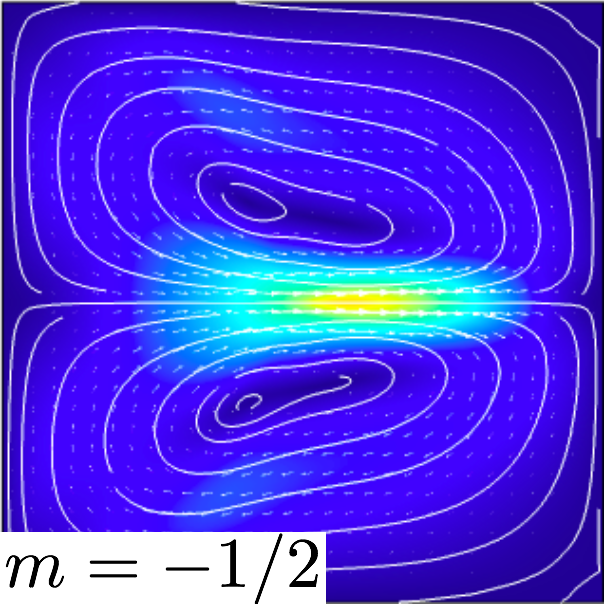}
\caption{\label{fig:mHalfV}}
\end{subfigure}
\begin{subfigure}{0.05\textwidth}
\includegraphics[width=\linewidth]{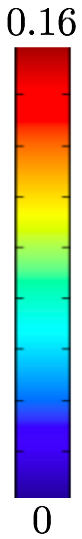}
\caption*{}
\end{subfigure}
\caption{Numerical results at $t=2\pi$ for single anchored disclinations with electric field applied in the horizontal direction. (\subref{fig:pHalfSurf})-(\subref{fig:mHalfSurf}) Plots of charge density within a square of dimensionless side length 2, centered at the disclination. (\subref{fig:pHalfV})-(\subref{fig:mHalfV}) Velocity across the entire cell for various disclinations. Color indicates velocity magnitude.}
\label{fig:ChargeVelocityMultipoles}
\end{figure*}

\begin{figure*}[h]
\centering
\begin{subfigure}{0.3\textwidth}
\includegraphics[width=\linewidth]{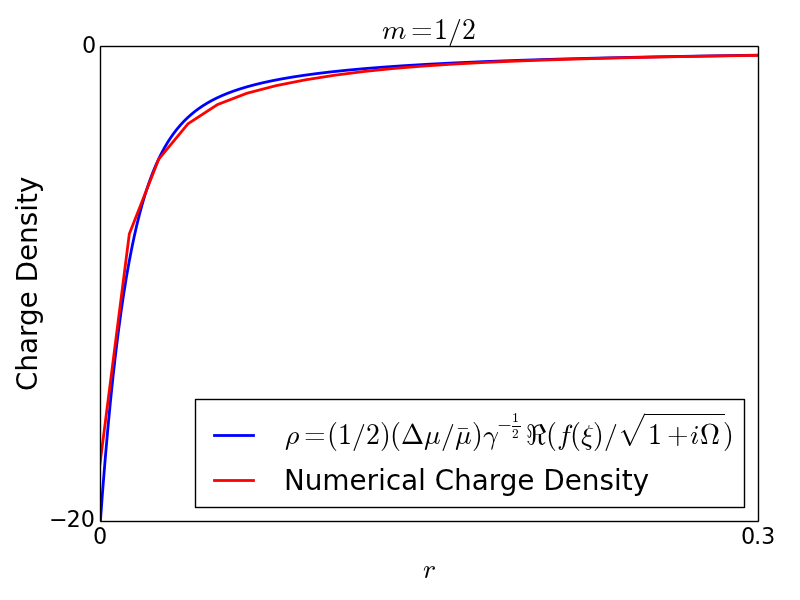}
\caption{\label{fig:pHalfCore}}
\end{subfigure}
\begin{subfigure}{0.3\textwidth}
\includegraphics[width=\linewidth]{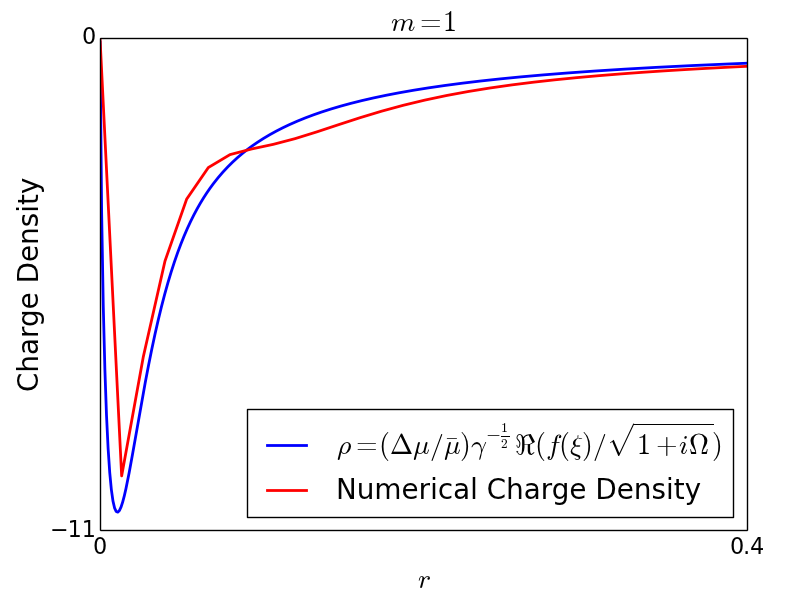}
\caption{\label{fig:p1Core}}
\end{subfigure}
\begin{subfigure}{0.3\textwidth}
\includegraphics[width=\linewidth]{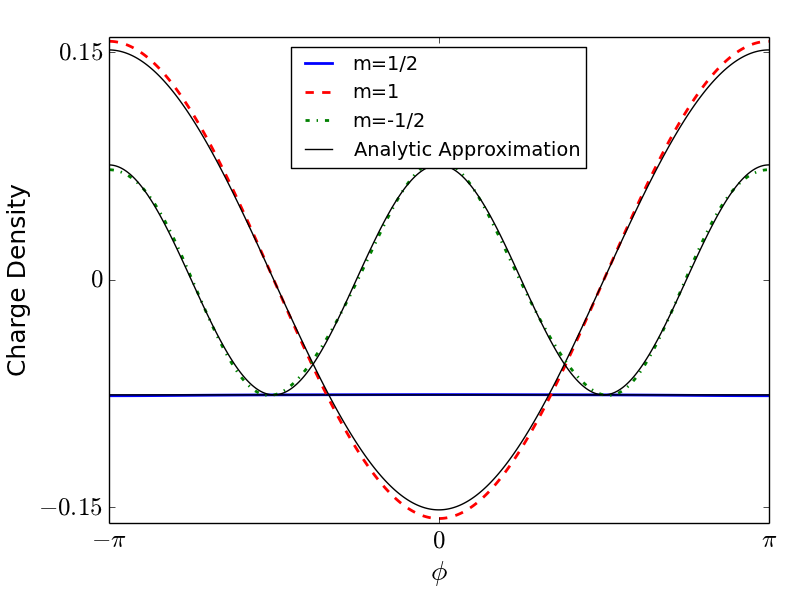}
\caption{\label{fig:AngularChargeComparison}}
\end{subfigure}
\caption{Numerical and analytical charge density results at $t=2\pi$ for single anchored disclinations. (\subref{fig:pHalfCore})-(\subref{fig:p1Core}) Numerical and analytical results along $\phi=0$ for $m=1/2$ and $m=1$. (\subref{fig:AngularChargeComparison}) Numerical and analytical results plotted as a function of angle at a distance $r=1.5$ from the disclination core.}
\label{fig:DisclinationComparison}
\end{figure*}

We compare these results to the finite element solutions to Eqs. (\ref{eq:concentration}) through (\ref{eq:leslie}). An advantage to the numerical model is that it does not contain the small parameter assumptions used in analytically obtaining Eq. (\ref{eq:rho_sol}).  Figure \ref{fig:ChargeVelocityMultipoles} shows the numerically-obtained charge densities and velocity fields for the cases $m=1/2, 1,$ and $-1/2$. Figure \ref{fig:DisclinationComparison} shows a comparison of the charge densities obtained numerically and our solutions to Eq. (\ref{eq:rho_scaled}) described above. We note that despite the simplifying assumptions, Eq. (\ref{eq:rho_sol}) agrees with the numerically-obtained charge density in both its angular and radial dependence.

In order to understand the flow structure shown in Fig. \ref{fig:ChargeVelocityMultipoles}, as well as the large distance behavior of the velocity under a body force that decays as a power law of distance away from the defect we solve the simpler problem of a Newtonian fluid,
\begin{equation}\label{eq:momentum_NS}
-\nabla p + \nabla^{2}\bv v - \rho\nabla\Phi=0;\quad \nabla\cdot\bv v=0,
\end{equation}
in a disk of radius 1, with $\bv v=0$ at $r=1$, and $\bv v$ finite for $r<1$. We consider only the part of the body force that does not time-average to zero, which far from the defect core is,
\begin{equation}
-\rho\nabla\Phi = -\left(\anisotropies\right)\frac{m \cos^2(t)}{(1+\Omega^2)}\frac{\cos[(2m-1)\phi]}{r}
\end{equation}

Taking the curl of Eq. (\ref{eq:momentum_NS}) and defining the stream function $\nabla\times (\psi_m \zhat) = -\bv v$ one obtains 
\begin{equation}\label{eq:vort2}
\nabla^{4}\psi_m(r,\phi,t) =\left(\anisotropies\right) \frac{m\cos^2(t)}{(1+\Omega^2)r^{2}}[m\sin(2m\phi)+(m-1)\sin(2(m-1)\phi)]
\end{equation}
Note that, as in the periodic case in Sec. \ref{sec:periodic}, the velocity is linear in the anisotropy difference, and in the high frequency limit, $\Omega\gg 1$, the systematic flow in the cell will disappear. Additionally, the angular dependence of the flow is set by the right-hand-side of Eq. (\ref{eq:vort2}). In particular, the angular flow structure will be a superposition of $2m$ and $2(m-1)$ harmonics.

For $m=1$ and $m=1/2$, the solutions to (\ref{eq:vort2}) satisfying the stated boundary conditions are
\begin{equation}\label{eq:psi_p1}
\psi_1(r,\phi,t)=\left(\anisotropies\right)\frac{\cos^2(t) r^{2}\sin(2\phi)}{16(1+\Omega^2)}\left[\frac12\left(1-r^2\right)+\log\left(r\right)\right]
\end{equation}
\begin{equation}\label{eq:psi_phalf}
\psi_{\frac12}(r,\phi,t)=\left(\anisotropies\right)\frac{\cos^2(t) r}{12(1+\Omega^2)}\left(r-1\right)^2\sin\phi
\end{equation}

To find solutions for $m \neq 1/2,1$, first note that the function
\begin{equation}
P_\alpha(r,\phi,t)=\left(\anisotropies\right)\frac{\cos^2(t) r^{2}\sin(\alpha\phi)}{(1+\Omega^2)(\alpha+2)\alpha(\alpha-2)}\left(\frac{(\alpha-2)}{2}r^\alpha-\frac{\alpha}{2}r^{\alpha-2}+1\right);\quad \alpha>2
\end{equation}
solves the equation
\begin{equation}\label{eq:vort_alpha}
\nabla^{4}P_\alpha(r,\phi,t) = \left(\anisotropies\right)\frac{\cos^2(t)\alpha}{2(1+\Omega^2)r^{2}}\sin(\alpha\phi)
\end{equation}
with $\partial P_\alpha/\partial r$ and $(1/r)(\partial P_\alpha/\partial \phi)$ finite for $r<1$, and
\begin{equation}
\left.\dbyd{P_\alpha}{r}\right|_{r=1}=\left.\frac{1}{r}\dbyd{P_\alpha}{\phi}\right|_{r=1}=0
\end{equation}

Using this solution we find the stream functions for $m=-1/2,-1,3/2$, and $2$,
\begin{equation}\label{eq:psi_mhalf}
\psi_{-\frac12}(r,\phi,t) =  -\frac12[\psi_{\frac12}(r,\phi,t)+P_3(r,\phi,t)]
\end{equation}
\begin{equation}
\psi_{-1}(r,\phi,t) = -\psi_1(r,\phi,t)-P_4(r,\phi,t)
\end{equation}
\begin{equation}
    \psi_{\frac32}(r,\phi,t)=\frac32[\psi_{\frac12}(r,\phi,t)+P_3(r,\phi,t)]
\end{equation}
\begin{equation}
    \psi_2(r,\phi,t) = 2[\psi_1(r,\phi,t)+P_4(r,\phi,t)]
\end{equation}
and the stream functions for all other values of $m$,
\begin{equation}
\psi_m(r,\phi,t) = m[P_{2m}(r,\phi,t)+P_{2(m-1)}(r,\phi,t)]
\end{equation}

Despite the assumption of a Newtonian fluid, the angular depencence of Eqs. (\ref{eq:psi_p1}) ($m=1$) and (\ref{eq:psi_phalf}) ($m = 1/2$) and (\ref{eq:psi_mhalf}) ($m=-1/2$) agrees with the flow fields shown in Fig. \ref{fig:ChargeVelocityMultipoles}. In addition, our results for a Newtonian fluid suggest that given the slow decay of the charge density created by a single disclination, the fluid velocity would diverge at large distances in an infinite domain. The bounded nature of our numerical results follow from the no slip boundary conditions at domain boundaries.

Having analyzed the electrokinetic behavior induced by a single disclination, we consider next configurations comprising a set of disclinations with total topological change zero, which may be used to engineer more complex flow structures. We follow the same procedure as described at the beginning of this section to find the equation for charge density to first order in $\Delta\mu/\bar\mu$ and $\Delta\epsilon/\bar\epsilon$ for $n$ disclinations:
\begin{equation}\label{eq:rho_mult}
\Omega\dbyd{\rho}{t} = \gamma\nabla^{2}\rho-\rho - \left(\anisotropies\right)\sum_{i=1}^n\frac{m_i\cos(2\theta(\bm r)-\phi_i)}{r_i}
\end{equation}
where $r_i=\sqrt{(x-x_i)^2+(y-y_i)^2}$ is the distance from disclination $i$ located at $(x_i,y_i)$, $\tan \phi_i = (y-y_i)/(x-x_i)$, $m_i$ is the topological charge of disclination $i$, and $\theta(\bv r) = \displaystyle\sum_{i=1}^n m_i\phi_i$ \footnote{In general, the director field created by a set of isolated dislocations is not the sum of the individual contributions except in the one elastic constant approximation to the Leslie-Ericksen model $K_{1} = K_{2} = K_{3}$.}. The solution to Eq. (\ref{eq:rho_mult}) far from disclination cores is,
\begin{equation}
\label{eq:charge_disclination}
\rho(\bm r,t) = -\left(\anisotropies\right)\frac{\cos(t-\delta)}{\sqrt{1+\Omega^2}}\sum_{i=1}^n\frac{m_i\cos(2\theta(\bm r)-\phi_i)}{r_i}, \quad\quad \tan\delta = \Omega,
\end{equation}

\begin{figure*}[h]
\centering
\begin{subfigure}{0.3\textwidth}
\includegraphics[width=\linewidth]{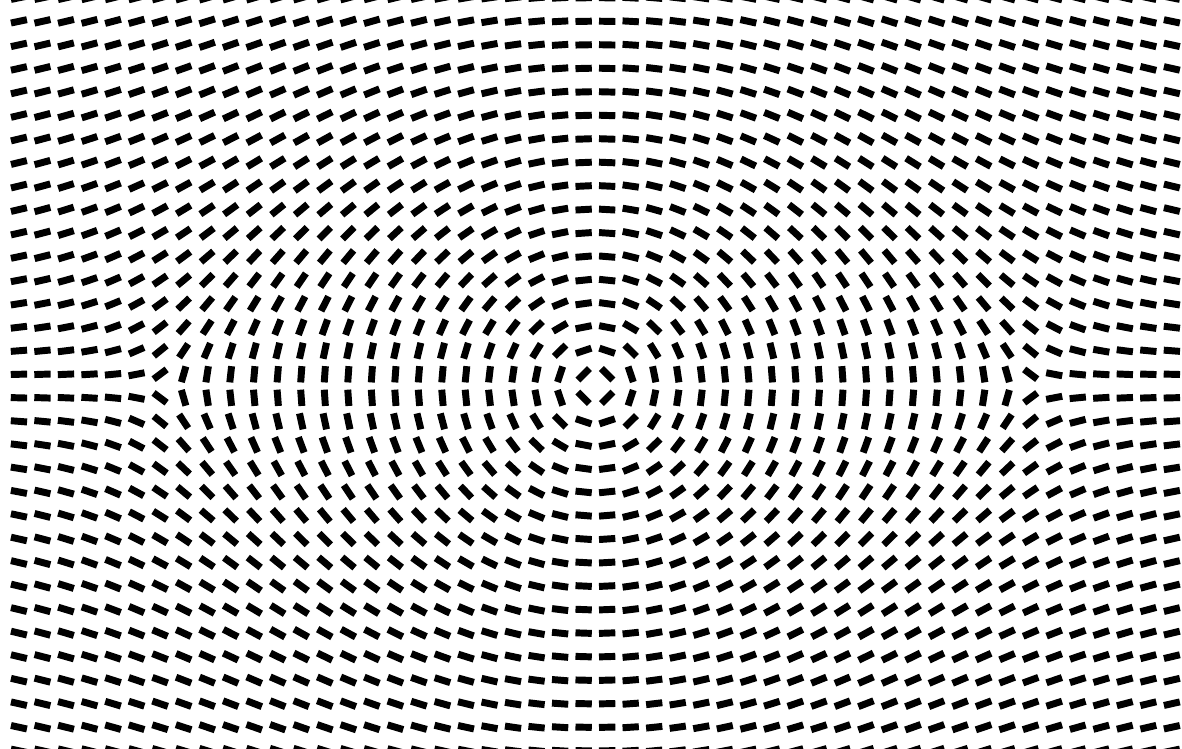}
\caption{\label{fig:3DefectPattern}}
\end{subfigure}
\begin{subfigure}{0.34\textwidth}
\includegraphics[width=\linewidth]{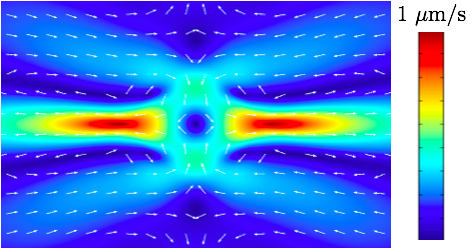}
\caption{\label{fig:3DefectNumeric}}
\end{subfigure}
\begin{subfigure}{0.34\textwidth}
\includegraphics[width=\linewidth]{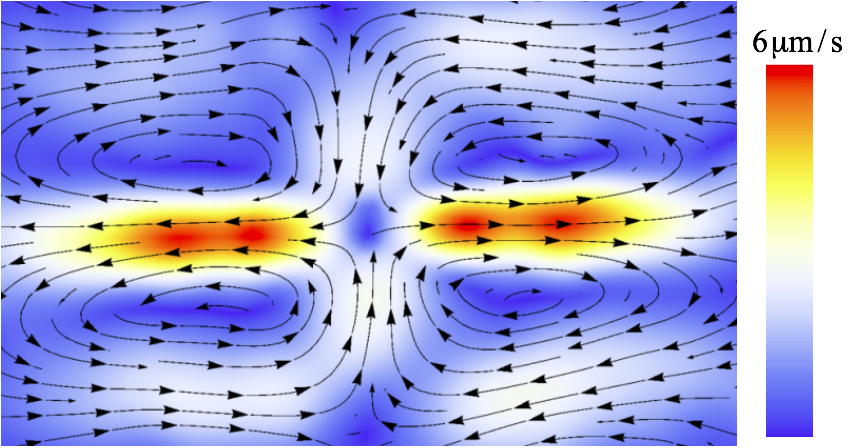}
\caption{\label{fig:3DefectExperiment}}
\end{subfigure}
\caption{(\subref{fig:3DefectPattern}) Pattern with three disclinations; two with charge -1/2 and one with charge +1. (\subref{fig:3DefectNumeric}) Time averaged velocity under AC field applied horizontally (\subref{fig:3DefectExperiment}) Time averaged experimental velocity for the same configuration.}
\label{fig:3DefectComparison}
\end{figure*}

\begin{figure}[h]
\begin{subfigure}{0.5\textwidth}
\includegraphics[width=\linewidth]{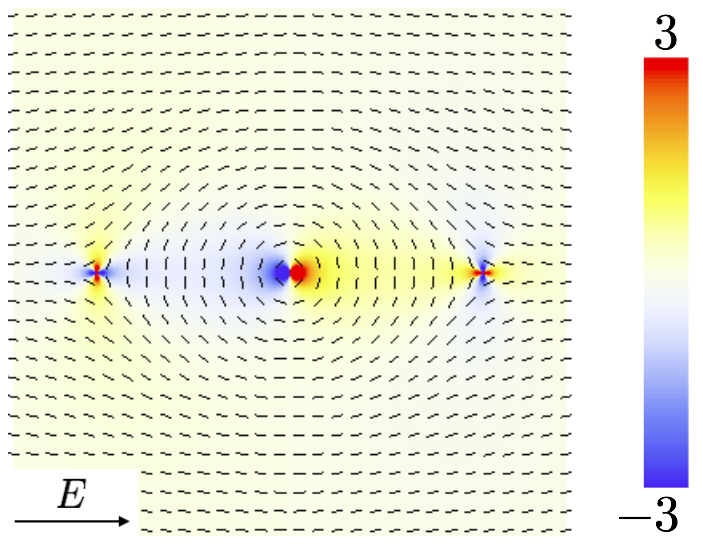}
\caption{\label{fig:3AnalyticCharge}}
\end{subfigure}
\begin{subfigure}{0.5\textwidth}
\includegraphics[width=\linewidth]{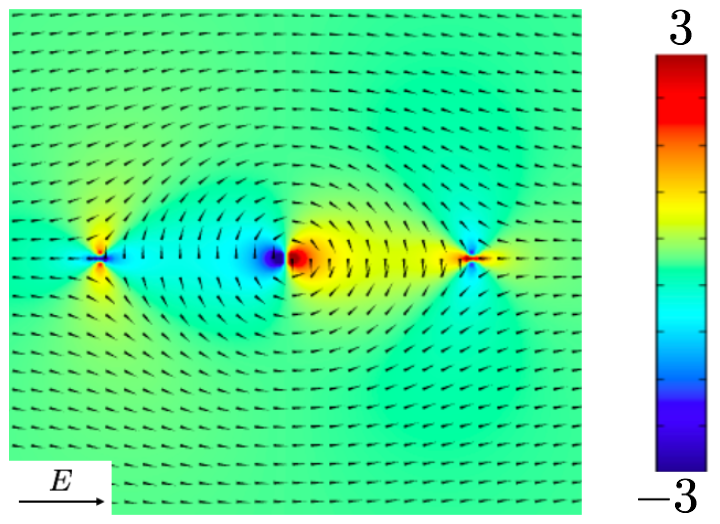}
\caption{\label{fig:3NumericCharge}}
\end{subfigure}
\caption{(\subref{fig:3AnalyticCharge}) Analytic charge density according to Eq. (\ref{eq:charge_disclination}) for a superposition of a (-1/2,1,-1/2) disclination triplet pattern expressed in dimensionless units at $t=2\pi$. (\subref{fig:3NumericCharge}) Numerical charge density in dimensionless units at $t=2\pi$ for the same parameters.}
\label{fig:3DefectCharges}
\end{figure}

Note that this solution reduces to Eq. (\ref{eq:rho_far}) when $n=1$. The solution is not quite a superposition of single-disclination charge densities, as the term $\cos(2\theta(\bm r)-\phi_i)$ contributes cross-terms to the sum in Eq. (\ref{eq:charge_disclination}). Figure \ref{fig:3DefectPattern} shows, for example, a configuration with three disclinations of topological charge (-1/2, 1, -1/2), studied experimentally by Peng, et al\cite{re:peng15}. In our numerical solution, Fig. \ref{fig:3DefectNumeric}, lengths are scaled by the defect separation. We use no-slip boundary conditions on the velocity in the far field, while in the experiments, the director pattern is only imposed in a small subdomain of the experimental cell.  Figure \ref{fig:3DefectExperiment} shows the experimental flow field, which is close both in structure and magnitude to that in Fig. \ref{fig:3DefectNumeric} which has been obtained numerically. Figure \ref{fig:3DefectCharges} compares Eq. (\ref{eq:charge_disclination}) to the numerical charge density obtained for the (-1/2, 1, -1/2) disclination pattern. 

We finally address the mechanism responsible for systematic fluid and particle motion in a configuration comprising a suspended spherical particle and an associated hedgehog defect when the AC electric field is {\em perpendicular} to the line joining the particle and the defect \cite{re:lazo13}. The net effect is a transverse mobility, different from the conventional electrophoresis case in which particle motion is parallel to the applied field. We do not directly use a suspended particle, rather we model the experimental configuration by two anchored disclinations with topological charges (+1) and (-1). The (+1) disclination represents the topological charge created by a spherical particle with homeotropic anchoring, and the (-1) defect models the accompanying hedgehog defect. Unlike the previous study in which lengths were scaled by the disclination separation, in this study we scale lengths by the radius of the particle represented by the (+1) disclination. The separation between the defects is taken equal to the distance between the center of the spherical particle and the defect in the experiments, estimated as $1.17$ by Poulin, et. al.\cite{re:poulin97}

\begin{figure*}[h]
\begin{subfigure}{0.31\textwidth}
\includegraphics[width=\linewidth]{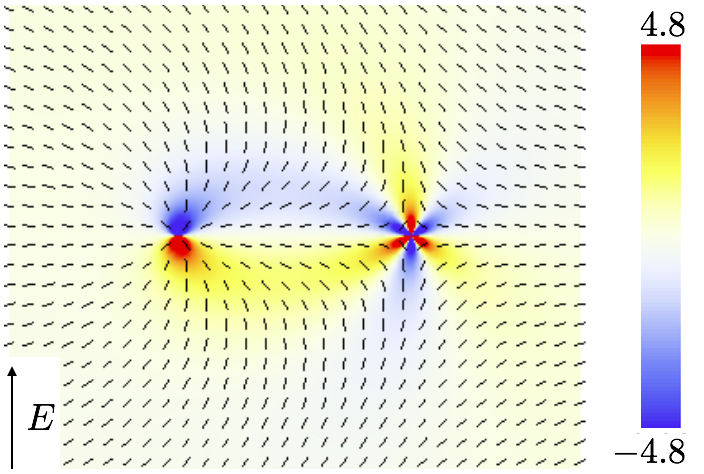}
\caption{\label{fig:AnalyticChargep1m1}}
\end{subfigure}
\begin{subfigure}{0.31\textwidth}
\includegraphics[width=\linewidth]{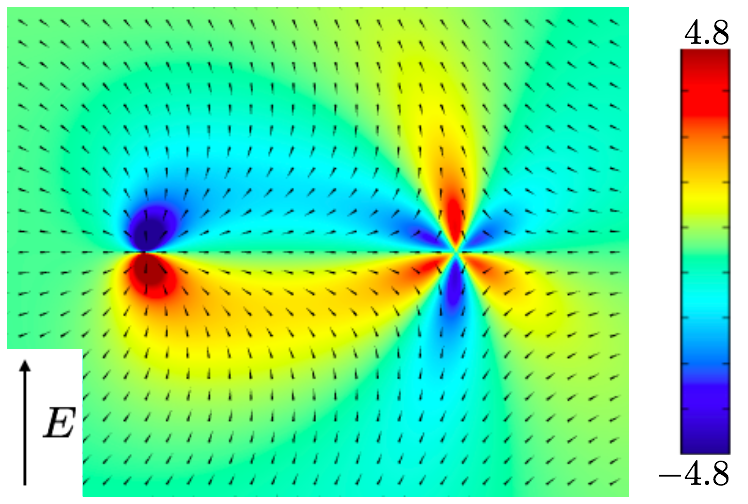}
\caption{\label{fig:PerpFieldCharge}}
\end{subfigure}
\begin{subfigure}{0.31\textwidth}
\includegraphics[width=\linewidth]{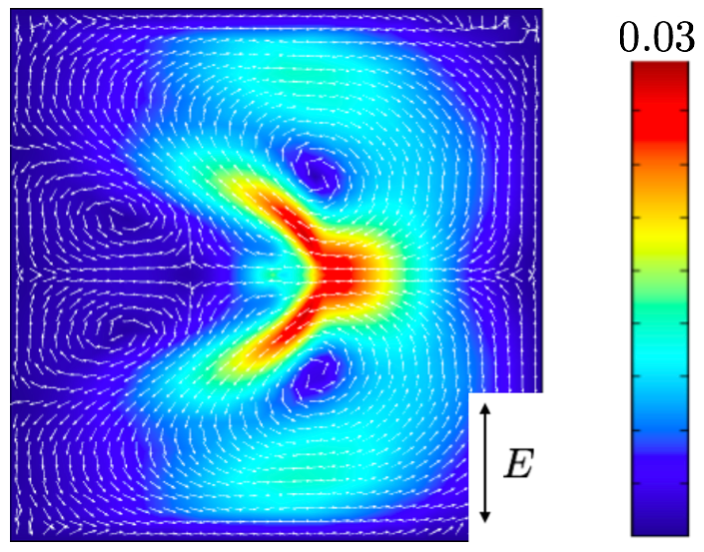}
\caption{\label{fig:PerpFieldVelocity}}
\end{subfigure}
\caption{Numerical and analytical results for two anchored disclinations with charges +1 and -1. The applied electric field in this case is applied in the vertical direction, rather than the horizontal direction. (\subref{fig:AnalyticChargep1m1}) Charge density according to Eq. (\ref{eq:charge_disclination_perp}), in dimensionless units at $t=2\pi$. (\subref{fig:PerpFieldCharge}) Numerical charge density in dimensionless units at $t=2\pi$. (\subref{fig:PerpFieldVelocity}) Numerical average velocity in dimensionless units. The asymmetry of the charge density leads to systematic flow in the horizontal direction.}
\label{fig:PerpField}
\end{figure*}

We follow the same steps as in deriving Eq. (\ref{eq:charge_disclination}) except with the applied potential in the vertical direction, and we obtain the charge density for this configuration as
\begin{equation}
\label{eq:charge_disclination_perp}
\rho(\bm r,t) = \left(\anisotropies\right)\frac{\cos( t-\delta)}{\sqrt{1+\Omega^2}}\sum_{i=1}^n\frac{m_i\sin(2\theta(\bm r)-\phi_i)}{r_i}, \quad\quad \tan\delta = \Omega.
\end{equation}
Figures \ref{fig:AnalyticChargep1m1} and \ref{fig:PerpFieldCharge} show the analytic and numerical solutions for the instantaneous charge density corresponding to the positive electric field (pointing upwards in the figure), whereas the average flow field is shown in Fig. \ref{fig:PerpFieldVelocity}. The asymmetry in sizes of the six charge density lobes of Fig. \ref{fig:PerpFieldCharge} leads to an equally asymmetric flow field. This flow field is quadratic in the field amplitude, and therefore does not change sign as the AC field does. Therefore, if the defects were free to move instead of anchored, this flow field would lead to systematic particle motion and to a transverse (off diagonal) mobility.

In summary, the transport model of Sec. \ref{sec:model} both qualitatively and quantitatively accounts for the main transport features in a nematic film with an imposed director field and subjected to an oscillatory, uniform, electrostatic field. The existence of anisotropy either in the liquid crystal molecular dielectric permittivity or in the mobility of ionic impurities in the fluid lead to spatial charge separation without requiring other solid surfaces as in conventional electrokinetic phenomena. For AC field frequencies that are low compared to the inverse charging time, the body force on the fluid is quadratic in the imposed field and leads to systematic electroosmotic flows in the cases considered. Both the spatial structure of the flows and the velocity amplitudes obtained for the case of periodically patterned director configurations as well as sets of isolated disclinations are in good agreement with the experiments. We have further elucidated the origin of a transverse mobility, in which defect motion results which is perpendicular to the imposed field. Of course, such a mobility is important in flow and particle control, and of potential experimental interest as it would allow a precise control of the motion of suspended particles.

\section{Acknowledgements}
We are indebted to Chenhui Peng and Oleg Lavrentovich for many stimulating discussions, and access to the experimental data sets. We alo thank Carme Calderer and Dmitry Golovaty for guidance in the model development. This research has been supported by the National Science Foundation under contract DMS 1435372, and the Minnesota Supercomputing Institute.

\appendix

\section{Numerical method and choice of parameters}
\label{sec:numeric}



Equations (\ref{eq:concentration}) through (\ref{eq:leslie}), together with the incompressibility condition, completely describe our system. We solve them numerically with the finite element commercial package COMSOL; the code developed is available for download\cite{re:conklin16}. Instead of Eq. (\ref{eq:directorDynamics}), our solution assumes the director $\bv{\hat n}$ is in elastic equilibrium, which implies\cite{re:degennes93} $\partial_j T_{ij}^e = -\partial_i f$. Thus following a procedure similar to Stark\cite{re:stark01}, we solve for $\tilde p = p+f$ rather than $p$ and write Eq. (\ref{eq:momentum_conservation}) as
\begin{equation}\label{eq:momentum_simplified}
0=-\dbyd{\tilde p}{x_i}+\dbyd{}{x_j}\tilde{T}_{ij}-\sum_{k=1}^N e z_kc_k\nabla\Phi
\end{equation}

The finite element representation chosen is Lagrange elements of order 2 for all variables except the pressure $\tilde p$, which uses Lagrange elements of order 1. All numerical solutions were obtained in a two-dimensional square domain $S_0$ with side length $\ell$. For periodic anchoring, we choose $\ell = 1$; for anchored disclinations, $\ell=15$. We define a coordinate system $(x,y)$, the origin of which is at the center of $S_0$. We specify the electrostatic potential at the boundaries $\Phi(x=\ell/2,y) = 0$ and $\Phi(x=-\ell/2,y) = -\ell \cos(t)$, so that the applied AC field is uniform and parallel to the $x$ axis.

\begin{figure}[h]
\centering
\begin{subfigure}{0.4\textwidth}
\includegraphics[width=\linewidth]{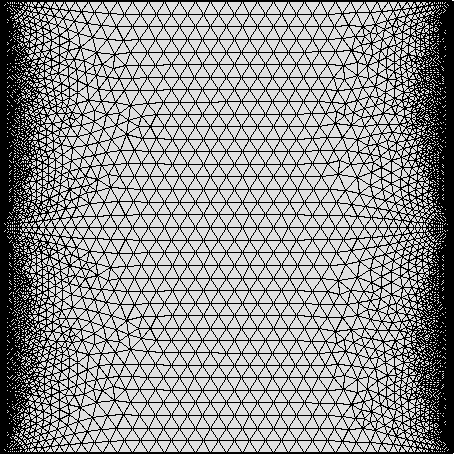}
\caption{\label{fig:PeriodicMesh}}
\end{subfigure}
\begin{subfigure}{0.4\textwidth}
\includegraphics[width=\linewidth]{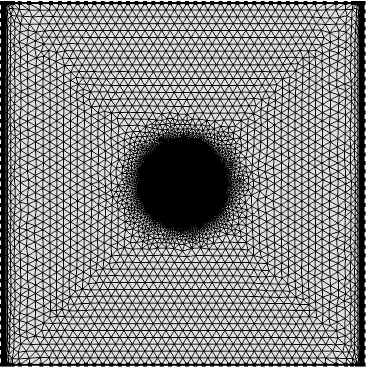}
\caption{\label{fig:DisclinationMesh}}
\end{subfigure}
\caption{(\subref{fig:PeriodicMesh}) Mesh used in the numerical solution in which $\hat{\bv n}$ is periodic. The mesh uses thin quadrilateral boundary-layer elements on the lateral walls and triangular elements throughout the remaining domain. (\subref{fig:DisclinationMesh}) Mesh used for the solution involving anchored disclinations. The mesh consists of triangular elements, more refined near the lateral walls. The central region is much more refined than the outer regions. }
\label{fig:Meshes}
\end{figure}

For periodic anchoring, periodic boundary conditions were used on the domain walls normal to the direction of the applied electric field ($y=\pm 1/2$). On $x=\pm 1/2$, no-slip boundary conditions were used for the velocity, and no flux for each of the ionic concentrations. Figure \ref{fig:PeriodicMesh} shows the mesh used for this case of periodic anchoring. Although COMSOL does have an option for periodic boundary conditions, to insure the mesh was periodic in the $y$ direction we created a mesh for half of the domain ($y>0$) and then reflected the mesh about the line $y=0$. The mesh consists of 9876 triangular elements of maximum linear size $3.5\times10^{-2}$ and minimum size $1.0\times 10^{-3}$. The mesh is finer near $x=\pm 1/2$ to resolve the boundary layer that forms there.

%
For anchored disclinations, we use no-slip boundary conditions for the velocity and zero flux boundary conditions for ion currents on all boundaries. The numerical implementation involves two meshes with different resolutions. Within the square $S_0$ we create a circle $C_0$ with radius $R=1.5$ in which the mesh is finer. Figure \ref{fig:DisclinationMesh} shows the mesh used for disclinations. $C_0$ contains 109,196 triangular elements, with maximum linear size $1.3\times 10^{-2}$, and minimum size $4.8\times 10^{-5}$. The remainder of $S_0$ contains 12,800 triangular elements with linear size between $1.2\times 10^{-3}$ and $1.0$, and 240 thin quadrilateral boundary layer elements near the domain boundaries at $x=\pm \ell/2$.

In order to study electrokinetic phenomena in realistic ranges of physical parameters, we have chosen a parameter set that corresponds to the experiments of Peng, et. al\cite{re:peng15}. While we have shown good agreement between the numerical and experimental results, our comparison with experiments is limited by a number of factors: First, our numerical model is solved on a square domain of size $\sim 1~\text{mm}^2$ or smaller, with the desired director pattern imposed throughout the cell, while the experiments are done in a chamber of dimensions 10 mm $\times$ 15 mm, with the desired pattern imposed on a $1~\text{mm}^2$ region and uniform alignment on the remainder of the cell surface. Second, our equations are solved in two dimensions, and while there is no evidence of the third dimension affecting the overall experimental behavior, one would expect a Pouiselle-type flow across the cell thickness that is unaccounted for in our model. Third, there is some non-uniformity in the electric field in the experimental cell that includes the fluid and the confining glass plates. This results in a decreased electric field on the sample which has not been quantified. In the experiments of Lazo, et. al\cite{re:lazo14}, the effective field was approximately 65 \% of the nominal field. The field amplitude used numerically is the value of the nominal electrostatic field from the experiments of Peng, et. al\cite{re:peng15}. Finally, the experiments use a mixture of MLC7026-000 and E7 liquid crystals at a ratio of 89.1:10.9\cite{re:lazo14}. The Leslie viscosities were not measured for this mixture, so viscosities in the numerical model were determined by averaging the viscosities of the two liquid crystals\cite{re:vanbrabant09,re:wang06} at the same ratio as above.

\begin{table}
\begin{center}\begin{tabular}{ccc}{\bf Parameter} & \bf Value & \bf Comment \\
$\omega$ & $10\pi$ rad/s & Applied field frequency\\
$E_0$ & 40 mV/$\mu$m & Applied field amplitude \\
$n_0$ & $10^{-19}~\text{m}^{-3}$ & Total ion concentration \\
$\bar\mu$ & $1.45\times 10^{-9}$ $\text{m}^2$/(Vs) & Average ion mobility \\
$\bar\epsilon$ & 6 & Average dielectric permittivity \\
$\Delta\mu/\bar\mu$ & 0.34 & Relative mobility anisotropy \\
$\Delta \epsilon/\bar\epsilon$ & 0 & Relative dielectric anisotropy \\
$\bar D$ & $k_B T\bar\mu/e = 3.69\times 10^{-11}$ $\text{m}^2$/s & Average ion diffusivity\\
$\bar\epsilon$ & 6 & Average dielectric constant \\
$\alpha_1$ & $-29$ mPa s & Leslie-Ericksen viscosities \\
$\alpha_2$ & $-173$ mPa s & Leslie-Ericksen viscosities \\
$\alpha_3$ & $-30$ mPa s & Leslie-Ericksen viscosities \\
$\alpha_4$ & $118$ mPa s & Leslie-Ericksen viscosities \\
$\alpha_5$ & $137$ mPa s & Leslie-Ericksen viscosities \\
$\alpha_6$ & $-66$ mPa s & Leslie-Ericksen viscosities
\end{tabular}
\caption{\label{tab:params} Physical constants used in numerical calculations. $e$ is the electron charge, $k_B$ is Boltzmann's constant, and $T = 295 K$, room temperature. The relative mobility and dielectric anisotropies are as listed except as noted in Fig. \ref{fig:AnisotropyComparison}. The characteristic length used is dependent on the director class studied.}
\end{center}
\end{table}

For periodic anchoring we choose the characteristic length $L$ to be equal to the wavelength of the director pattern, $L=2\pi/q = 160~\mu\text{m}$. For anchored disclinations under a field parallel to $\bv{\hat{n}}$, we choose $L=80~\mu\text{m}$, the separation between disclinations\cite{re:peng15}. For our study of disclinations under a transverse field, we choose $L=25~\mu\text{m}$ corresponding to the radius of the particle modeled by the $(+1)$ disclination. The remaining parameters are the same for both pattern classes, and are listed in Table \ref{tab:params}.

\section{Effect of concentration gradients on charge density}
\label{sec:term_neglection}

We wish to show that spatial variations in $C = c_1 + c_2$ contribute negligibly to the charge density $\rho_1$, given that $Y^2/(4\gamma\sqrt{1+\Omega^2})\ll1$. To begin we Fourier transform Eqs. (\ref{eq:C_n}) and (\ref{eq:rho_n}) in space,
\begin{equation}
\hat b_n(\bv q) = -\frac{Y^2 i q_x}{2(\gamma q^2+i n \Omega)}[\hat a_{n-1}(\bv q)+\hat a_{n+1}(\bv q)]
\end{equation}
\begin{equation}\label{eq:rho_fourier}
(\Omega n i+\gamma q^2+1)\hat{a}_n(\bv q) = -\left(\frac{q_x Y}{2}\right)^2\left[\frac{\hat{a}_{n-2}(\bv q)+\hat{a}_{n}(\bv q)}{\gamma q^2+i(n-1)\Omega}+\frac{\hat{a}_n(\bv q)+\hat{a}_{n+2}(\bv q)}{\gamma q^2+i(n+1)\Omega}\right]-F(\bv q)(\delta_{n,1}+\delta_{n,-1})
\end{equation}
where
\begin{equation}
F(\bv q) = \iint e^{-i\bv q\cdot \bv x}  d^2 x \left(\anisotropies\right)\frac{m \cos[(2m-1)\phi]}{2r}
\end{equation}

Note that 
\begin{equation}\label{eq:inequality}
\left|\left(\frac{q_x Y}{2}\right)^2\frac{1}{[\gamma q^2+i(n\pm 1)\Omega][\gamma q^2+1+\Omega n i]}\right| \leq \frac{Y^2}{4\gamma\sqrt{1+\Omega^2}}\ll 1
\end{equation}
Which implies 
$$\left|\left(\frac{q_x Y}{2}\right)^2\frac{\hat{a}_n}{\gamma q^2+i(n\pm 1)\Omega}\right| \ll |(\Omega n i + \gamma q^2+1)\hat{a}_n|$$
so we may approximate Eq. (\ref{eq:rho_fourier}) as
\begin{equation}\label{eq:rho_fourier2}
(\Omega n i+\gamma q^2+1)\hat{a}_n(\bv q) = -\left(\frac{q_x Y}{2}\right)^2\left[\frac{\hat{a}_{n-2}(\bv q)}{\gamma q^2+i(n-1)\Omega}+\frac{\hat{a}_{n+2}(\bv q)}{\gamma q^2+i(n+1)\Omega}\right]-F(\bv q)(\delta_{n,1}+\delta_{n,-1})
\end{equation}

We assume $\rho$ is continuous and differentiable in time and space. Therefore $|a_{n}|\rightarrow 0$ as $|n|\rightarrow\infty$. We use this assumption to show $|\hat{a}_3|\ll |\hat{a}_1|$. First suppose $|\hat{a}_{n-2}|\lesssim|\hat{a}_n|$ for $n\geq3$. Inequality (\ref{eq:inequality}) along with Eq. (\ref{eq:rho_fourier2}) then implies $|\hat{a}_{n}| \ll |\hat{a}_{n+2}|$. Thus if $|\hat{a}_1|\lesssim|\hat{a}_3|$, then by induction $|\hat a_n|$ does not approach zero as $n\rightarrow\infty$, which contradicts our assumption that $\rho$ is continuous and differentiable. Therefore $|\hat{a}_3|\ll|\hat a_1|$.

Equation (\ref{eq:rho_fourier2}) for $n=1$ is
\begin{equation}\label{eq:rho_1}
(\gamma q^2+1+\Omega i)\hat{a}_1(\bv q) =  -\left(\frac{q_x Y}{2}\right)^2\frac{\hat{a}_{-1}(\bv q)}{\gamma q^2}-\left(\frac{q_x Y}{2}\right)^2\frac{\hat{a}_{3}(\bv q)}{\gamma q^2+2\Omega i}-F(\bv q)
\end{equation}
Inequality (\ref{eq:inequality}) and the fact that $|\hat{a}_3|\ll |\hat a_1|$ imply the second term on the right-hand-side of Eq. (\ref{eq:rho_1}) is negligible relative to the left-hand-side. Additionally,
$$\frac{\left|\left(\frac{q_x Y}{2}\right)^2\frac{\hat{a}_{-1}}{\gamma q^2}\right|}{|(\gamma q^2+1+\Omega i)\hat{a}_1 |} = \left(\frac{q_x Y}{2}\right)^2\frac{1}{\gamma q^2\sqrt{(\gamma q^2+1)^2+\Omega^2}}\leq \frac{Y^2}{4\gamma\sqrt{1+\Omega^2}}\ll 1$$
Thus the first term in Eq. (\ref{eq:rho_1}) is negligible, and we may approximate Eq. (\ref{eq:rho_1}) as
\begin{equation}
(\gamma q^2+1+\Omega i)\hat{a}_1(\bv q) =-F(\bv q)
\end{equation}
which is the Fourier transform of Eq. (\ref{eq:rho_simp}).

\bibliography{patterned_lc}

\providecommand*{\mcitethebibliography}{\thebibliography}
\csname @ifundefined\endcsname{endmcitethebibliography}
{\let\endmcitethebibliography\endthebibliography}{}
\begin{mcitethebibliography}{26}
\providecommand*{\natexlab}[1]{#1}
\providecommand*{\mciteSetBstSublistMode}[1]{}
\providecommand*{\mciteSetBstMaxWidthForm}[2]{}
\providecommand*{\mciteBstWouldAddEndPuncttrue}
  {\def\EndOfBibitem{\unskip.}}
\providecommand*{\mciteBstWouldAddEndPunctfalse}
  {\let\EndOfBibitem\relax}
\providecommand*{\mciteSetBstMidEndSepPunct}[3]{}
\providecommand*{\mciteSetBstSublistLabelBeginEnd}[3]{}
\providecommand*{\EndOfBibitem}{}
\mciteSetBstSublistMode{f}
\mciteSetBstMaxWidthForm{subitem}
{(\emph{\alph{mcitesubitemcount}})}
\mciteSetBstSublistLabelBeginEnd{\mcitemaxwidthsubitemform\space}
{\relax}{\relax}

\bibitem[Morgan and Green(2003)]{re:morgan03}
H.~Morgan and N.~Green, \emph{AC electrokinetics: colloids and nanoparticles},
  Research Studies Press Ltd., Baldock, UK, 2003\relax
\mciteBstWouldAddEndPuncttrue
\mciteSetBstMidEndSepPunct{\mcitedefaultmidpunct}
{\mcitedefaultendpunct}{\mcitedefaultseppunct}\relax
\EndOfBibitem
\bibitem[Squires and Bazant(2004)]{re:squires04}
T.~Squires and M.~Bazant, \emph{J. Fluid Mech.}, 2004, \textbf{509},
  217--252\relax
\mciteBstWouldAddEndPuncttrue
\mciteSetBstMidEndSepPunct{\mcitedefaultmidpunct}
{\mcitedefaultendpunct}{\mcitedefaultseppunct}\relax
\EndOfBibitem
\bibitem[Bazant \emph{et~al.}(2009)Bazant, Kilic, Storey, and
  Ajdari]{re:bazant09}
M.~Z. Bazant, M.~S. Kilic, B.~D. Storey and A.~Ajdari, \emph{Advances in
  Colloid and Interface Science}, 2009, \textbf{152}, 48 -- 88\relax
\mciteBstWouldAddEndPuncttrue
\mciteSetBstMidEndSepPunct{\mcitedefaultmidpunct}
{\mcitedefaultendpunct}{\mcitedefaultseppunct}\relax
\EndOfBibitem
\bibitem[Stark(2001)]{re:stark01}
H.~Stark, \emph{Phys. Rep.}, 2001, \textbf{351}, 387--474\relax
\mciteBstWouldAddEndPuncttrue
\mciteSetBstMidEndSepPunct{\mcitedefaultmidpunct}
{\mcitedefaultendpunct}{\mcitedefaultseppunct}\relax
\EndOfBibitem
\bibitem[Pishnyak \emph{et~al.}(2007)Pishnyak, Tang, Kelly, Shiyanovskii, and
  Lavrentovich]{re:pishnyak07}
O.~Pishnyak, S.~Tang, J.~Kelly, S.~Shiyanovskii and O.~Lavrentovich,
  \emph{Phys. Rev. Lett.}, 2007, \textbf{99}, 127802\relax
\mciteBstWouldAddEndPuncttrue
\mciteSetBstMidEndSepPunct{\mcitedefaultmidpunct}
{\mcitedefaultendpunct}{\mcitedefaultseppunct}\relax
\EndOfBibitem
\bibitem[{Turiv} \emph{et~al.}(2013){Turiv}, {Lazo}, {Brodin}, {Lev},
  {Reiffenrath}, {Nazarenko}, and {Lavrentovich}]{re:turiv13}
T.~{Turiv}, I.~{Lazo}, A.~{Brodin}, B.~I. {Lev}, V.~{Reiffenrath}, V.~G.
  {Nazarenko} and O.~D. {Lavrentovich}, \emph{Science}, 2013, \textbf{342},
  1351--1354\relax
\mciteBstWouldAddEndPuncttrue
\mciteSetBstMidEndSepPunct{\mcitedefaultmidpunct}
{\mcitedefaultendpunct}{\mcitedefaultseppunct}\relax
\EndOfBibitem
\bibitem[Poulin \emph{et~al.}(1997)Poulin, Stark, Lubensky, and
  Weitz]{re:poulin97}
P.~Poulin, H.~Stark, T.~Lubensky and D.~Weitz, \emph{Science}, 1997,
  \textbf{275}, 1770--1773\relax
\mciteBstWouldAddEndPuncttrue
\mciteSetBstMidEndSepPunct{\mcitedefaultmidpunct}
{\mcitedefaultendpunct}{\mcitedefaultseppunct}\relax
\EndOfBibitem
\bibitem[Araki and Tanaka(2006)]{re:araki06}
T.~Araki and H.~Tanaka, \emph{Phys. Rev. Lett.}, 2006, \textbf{97},
  127801\relax
\mciteBstWouldAddEndPuncttrue
\mciteSetBstMidEndSepPunct{\mcitedefaultmidpunct}
{\mcitedefaultendpunct}{\mcitedefaultseppunct}\relax
\EndOfBibitem
\bibitem[Lavrentovich \emph{et~al.}(2010)Lavrentovich, Lazo, and
  Pishnyak]{re:lavrentovich10}
O.~Lavrentovich, I.~Lazo and O.~Pishnyak, \emph{Nature}, 2010, \textbf{467},
  947--950\relax
\mciteBstWouldAddEndPuncttrue
\mciteSetBstMidEndSepPunct{\mcitedefaultmidpunct}
{\mcitedefaultendpunct}{\mcitedefaultseppunct}\relax
\EndOfBibitem
\bibitem[{Hern\'andez}-Navarro \emph{et~al.}(2014){Hern\'andez}-Navarro,
  Tierno, Farrera, Ignes-Mullol, and {Sagu\'es}]{re:hernandez-navarro14}
S.~{Hern\'andez}-Navarro, P.~Tierno, J.~Farrera, J.~Ignes-Mullol and
  F.~{Sagu\'es}, \emph{Angewandte Chemie-International Edition}, 2014,
  \textbf{53}, 10696--106700\relax
\mciteBstWouldAddEndPuncttrue
\mciteSetBstMidEndSepPunct{\mcitedefaultmidpunct}
{\mcitedefaultendpunct}{\mcitedefaultseppunct}\relax
\EndOfBibitem
\bibitem[Lavrentovich(2015)]{re:lavrentovich15}
O.~Lavrentovich, \emph{arXiv:1512.04398}, 2015\relax
\mciteBstWouldAddEndPuncttrue
\mciteSetBstMidEndSepPunct{\mcitedefaultmidpunct}
{\mcitedefaultendpunct}{\mcitedefaultseppunct}\relax
\EndOfBibitem
\bibitem[Peng \emph{et~al.}(2015)Peng, Guo, Conklin, Vi\~nals, Shiyanovskii,
  Wei, and Lavrentovich]{re:peng15}
C.~Peng, Y.~Guo, C.~Conklin, J.~Vi\~nals, S.~V. Shiyanovskii, Q.-H. Wei and
  O.~D. Lavrentovich, \emph{Phys. Rev. E}, 2015, \textbf{92}, 052502\relax
\mciteBstWouldAddEndPuncttrue
\mciteSetBstMidEndSepPunct{\mcitedefaultmidpunct}
{\mcitedefaultendpunct}{\mcitedefaultseppunct}\relax
\EndOfBibitem
\bibitem[Kaiser \emph{et~al.}(1992)Kaiser, Pesch, and Bodenschatz]{re:kaiser92}
M.~Kaiser, W.~Pesch and E.~Bodenschatz, \emph{Physica D}, 1992, \textbf{59},
  320\relax
\mciteBstWouldAddEndPuncttrue
\mciteSetBstMidEndSepPunct{\mcitedefaultmidpunct}
{\mcitedefaultendpunct}{\mcitedefaultseppunct}\relax
\EndOfBibitem
\bibitem[Lazo and Lavrentovich(2013)]{re:lazo13}
I.~Lazo and O.~D. Lavrentovich, \emph{Phil. Trans. R. Soc. London A:
  Mathematical, Physical and Engineering Sciences}, 2013, \textbf{371},
  year\relax
\mciteBstWouldAddEndPuncttrue
\mciteSetBstMidEndSepPunct{\mcitedefaultmidpunct}
{\mcitedefaultendpunct}{\mcitedefaultseppunct}\relax
\EndOfBibitem
\bibitem[Lazo \emph{et~al.}(2014)Lazo, Peng, Xiang, Shiyanovskii, and
  Lavrentovich]{re:lazo14}
I.~Lazo, C.~Peng, J.~Xiang, S.~Shiyanovskii and O.~Lavrentovich, \emph{Nature
  Comm.}, 2014, \textbf{5}, 5033\relax
\mciteBstWouldAddEndPuncttrue
\mciteSetBstMidEndSepPunct{\mcitedefaultmidpunct}
{\mcitedefaultendpunct}{\mcitedefaultseppunct}\relax
\EndOfBibitem
\bibitem[Helfrich(1969)]{re:helfrich69}
W.~Helfrich, \emph{J. Chem. Phys.}, 1969, \textbf{51}, 4092\relax
\mciteBstWouldAddEndPuncttrue
\mciteSetBstMidEndSepPunct{\mcitedefaultmidpunct}
{\mcitedefaultendpunct}{\mcitedefaultseppunct}\relax
\EndOfBibitem
\bibitem[de~Gennes and Prost(1993)]{re:degennes93}
P.~de~Gennes and J.~Prost, \emph{The Physics of Liquid Crystals}, Clarendon,
  Oxford, 1993\relax
\mciteBstWouldAddEndPuncttrue
\mciteSetBstMidEndSepPunct{\mcitedefaultmidpunct}
{\mcitedefaultendpunct}{\mcitedefaultseppunct}\relax
\EndOfBibitem
\bibitem[Stewart(2004)]{re:stewart04}
I.~Stewart, \emph{The static and dynamic continuum theory of liquid crystals},
  Taylor and Francis, New York, 2004\relax
\mciteBstWouldAddEndPuncttrue
\mciteSetBstMidEndSepPunct{\mcitedefaultmidpunct}
{\mcitedefaultendpunct}{\mcitedefaultseppunct}\relax
\EndOfBibitem
\bibitem[de~Groot and Mazur(1984)]{re:degroot84}
S.~de~Groot and P.~Mazur, \emph{Non-equilibrium Thermodynamics}, Dover, New
  York, 1984\relax
\mciteBstWouldAddEndPuncttrue
\mciteSetBstMidEndSepPunct{\mcitedefaultmidpunct}
{\mcitedefaultendpunct}{\mcitedefaultseppunct}\relax
\EndOfBibitem
\bibitem[Kleman and Lavrentovich(2003)]{re:kleman03}
M.~Kleman and O.~Lavrentovich, \emph{Soft Matter Physics: An Introduction},
  Springer, New York, 2003\relax
\mciteBstWouldAddEndPuncttrue
\mciteSetBstMidEndSepPunct{\mcitedefaultmidpunct}
{\mcitedefaultendpunct}{\mcitedefaultseppunct}\relax
\EndOfBibitem
\bibitem[Calderer \emph{et~al.}(2016)Calderer, Golovaty, Lavrentovich, and
  Walkington]{re:calderer16}
M.~Calderer, D.~Golovaty, O.~Lavrentovich and N.~Walkington,
  \emph{arXiv:1601.02318}, 2016\relax
\mciteBstWouldAddEndPuncttrue
\mciteSetBstMidEndSepPunct{\mcitedefaultmidpunct}
{\mcitedefaultendpunct}{\mcitedefaultseppunct}\relax
\EndOfBibitem
\bibitem[Rien{\"a}cker \emph{et~al.}(2002)Rien{\"a}cker, Kr{\"o}ger, and
  Hess]{re:rienacker02}
G.~Rien{\"a}cker, M.~Kr{\"o}ger and S.~Hess, \emph{Phys. Rev. E}, 2002,
  \textbf{66}, 040702\relax
\mciteBstWouldAddEndPuncttrue
\mciteSetBstMidEndSepPunct{\mcitedefaultmidpunct}
{\mcitedefaultendpunct}{\mcitedefaultseppunct}\relax
\EndOfBibitem
\bibitem[Abramowitz and Stegun(1972)]{re:abramowitz72}
\emph{{Handbook of Mathematical Functions with Formulas, Graphs, and
  Mathematical Tables}}, ed. M.~Abramowitz and I.~A. Stegun, U.S. Dept. of
  Commerce, 1972, p. 498\relax
\mciteBstWouldAddEndPuncttrue
\mciteSetBstMidEndSepPunct{\mcitedefaultmidpunct}
{\mcitedefaultendpunct}{\mcitedefaultseppunct}\relax
\EndOfBibitem
\bibitem[Conklin(2016)]{re:conklin16}
C.~Conklin, \emph{\uppercase{COMSOL} code. Open Science Framework.
  \uppercase{DOI}:10.17605/OSF.IO/UBQS9}, 2016\relax
\mciteBstWouldAddEndPuncttrue
\mciteSetBstMidEndSepPunct{\mcitedefaultmidpunct}
{\mcitedefaultendpunct}{\mcitedefaultseppunct}\relax
\EndOfBibitem
\bibitem[Vanbrabant \emph{et~al.}(2009)Vanbrabant, Beeckman, Neyts, James, and
  Fernandez]{re:vanbrabant09}
J.~Vanbrabant, J.~Beeckman, K.~Neyts, R.~James and F.~Fernandez, \emph{Appl.
  Phys. Lett.}, 2009, \textbf{95}, 193502\relax
\mciteBstWouldAddEndPuncttrue
\mciteSetBstMidEndSepPunct{\mcitedefaultmidpunct}
{\mcitedefaultendpunct}{\mcitedefaultseppunct}\relax
\EndOfBibitem
\bibitem[Wang \emph{et~al.}(2006)Wang, Wu, Gauza, Wu, and Wu]{re:wang06}
H.~Wang, T.~Wu, S.~Gauza, J.~Wu and S.-T. Wu, \emph{Liquid Crystals}, 2006,
  \textbf{33}, 91\relax
\mciteBstWouldAddEndPuncttrue
\mciteSetBstMidEndSepPunct{\mcitedefaultmidpunct}
{\mcitedefaultendpunct}{\mcitedefaultseppunct}\relax
\EndOfBibitem
\end{mcitethebibliography}
\bibliographystyle{rsc} 

\end{document}